\documentclass[prd,showpacs,twocolumn,superscriptaddress,floatfix,longbibliography,nofootinbib]{revtex4-1}
\usepackage{graphicx,amsfonts,amssymb,amsmath,xspace,dsfont}
\usepackage[colorlinks=true,citecolor=green,linkcolor=red,urlcolor=blue]{hyperref}
\usepackage{subfigure}
\usepackage{blindtext}

\usepackage{color}
\definecolor{g}{rgb}{.1,0.4,.1} 
\definecolor{b}{rgb}{0,0.2,1}
\definecolor{rouge}{rgb}{0.82,0.,0.}
\definecolor{vert}{rgb}{0.,0.82,0.}
\definecolor{orange}{rgb}{1,0.5,0.}
\definecolor{bleu}{rgb}{0.,0.,0.82}
\definecolor{m}{rgb}{0.82,0.,0.82}
\definecolor{vert2}{rgb}{0.,0.5,0.}
\definecolor{rougeclair}{rgb}{1.0,0.7,0.7}

\newcommand{\beq}{\begin{equation}}
\newcommand{\be}{\begin{equation}}
\newcommand{\beqn}{\begin{eqnarray}}
\newcommand{\eeq}{\end{equation}}
\newcommand{\ee}{\end{equation}}
\newcommand{\eeqn}{\end{eqnarray}}

\newcommand{\ep}{{\epsilon}}

\newcommand{\bem}{\begin{pmatrix}}
\newcommand{\eem}{\end{pmatrix}}

\newcommand{\e}{\textrm{e}}
\newcommand{\im}{\textrm{i}}

\newlength{\ldag}
\begin{document}

\title{Landau levels in quasicrystals}

\author{Jean-No\"el Fuchs}
\email{fuchs@lptmc.jussieu.fr}
\affiliation{Sorbonne Universit\'e, CNRS, Laboratoire de Physique Th\'eorique de la Mati\`ere Condens\'ee, LPTMC, 75005 Paris, France}
\affiliation{Laboratoire de Physique des Solides, CNRS, Universit\'e Paris-Sud, Universit\'e Paris-Saclay, 91405 Orsay, France}

\author{R\'emy Mosseri}
\email{remy.mosseri@upmc.fr}
\affiliation{Sorbonne Universit\'e, CNRS, Laboratoire de Physique Th\'eorique de la Mati\`ere Condens\'ee, LPTMC, 75005 Paris, France}
\affiliation{Laboratory for Quantum Engineering and Micro-Nano Energy Technology and School of Physics and Optoelectronics, Xiangtan University, Hunan 411105, People's Republic of China}

\author{Julien Vidal}
\email{vidal@lptmc.jussieu.fr}
\affiliation{Sorbonne Universit\'e, CNRS, Laboratoire de Physique Th\'eorique de la Mati\`ere Condens\'ee, LPTMC, 75005 Paris, France}

\date{\today}

\begin{abstract}
Two-dimensional tight-binding models for quasicrystals made of plaquettes with commensurate areas are considered. Their energy spectrum is computed as a function of an applied perpendicular magnetic field. Landau levels are found to emerge near band edges in the zero-field limit. Their existence is related to an effective zero-field dispersion relation valid in the continuum limit. For quasicrystals studied here, an underlying periodic crystal exists and provides a natural interpretation to this dispersion relation. In addition to the slope (effective mass) of Landau levels, we also study their width as a function of the magnetic flux and identify two fundamental broadening mechanisms: (i) tunneling between closed cyclotron orbits and (ii) individual energy displacement of states within a Landau level. Interestingly, the typical broadening of the Landau levels is found to behave algebraically with the magnetic field with a nonuniversal exponent.  
\end{abstract}


\maketitle

%
\section{Introduction}
%

Due to translational invariance either in the continuum or in a periodic crystal, electrons are characterized by a dispersion relation giving the energy as a function of a conserved wave vector. In the presence of a magnetic field $B$ perpendicular to a two-dimensional (2D) system, the electrons classically follow iso-energy trajectories called cyclotron orbits. If the dispersion relation is such that it admits closed cyclotron orbits, the latter are quantized into discrete energy levels \cite{Onsager52}. The quantized cyclotron orbits are known as Landau levels (LLs) and have a macroscopic degeneracy $B \mathcal{A}$ (in units of the flux quantum) reflecting the translational invariance, where $\mathcal{A}$ is the sample area \cite{Landau30}. 

This macroscopic degeneracy can be lifted by different mechanisms leading to a broadening of LLs into Landau bands. 
First, a weak disordered potential of standard deviation $v$ produces a LL width that behaves as $v\sqrt{B}$, as shown by Ando \cite{Ando74b}. Second, the degeneracy can be lifted by applying a periodic (cosine) potential of strength $V_0$ and period $a$. Using first-order degenerate perturbation theory, Rauh \cite{Rauh75} has shown that the LL width behaves as $V_0 {\mathrm e}^{-\pi /(2Ba^2)}$ (see Appendix~\ref{app:rauh}). A third situation occurs in the case of a periodic crystal described by a tight-binding model. There, it is known from the Hofstadter butterfly \cite{Hofstadter76}, i.e., the energy spectrum versus magnetic field, that LLs are broadened by the presence of the underlying periodic potential. In an elegant semiclassical analysis, Wilkinson \cite{Wilkinson84} has shown that this broadening is due to tunneling between quantized cyclotron orbits (see Appendix~\ref{app:wilkinson}). The width behaves as $\sqrt{f}\:{\mathrm e}^{-c/f}$, where the coefficient $c$ can be obtained from a semiclassical Wentzel-Kramers-Brillouin (WKB) calculation, and $f=B\mathcal{A}_p$ is the dimensionless magnetic flux per plaquette of area $\mathcal{A}_p$.

We have recently computed the Hofstadter butterfly for a quasiperiodic tiling (Rauzy tiling) and also for a random tiling \cite{Fuchs16}. In both cases, LLs have been found near band edges and at low magnetic field. From the above discussion, it may appear as a surprise that LLs are also present in cases where translation invariance is absent and no dispersion relation \textit{stricto sensu} exists for $B=0$. It is well known that LLs survive a small amount of disorder, but here the nonperiodicity is strong. In the present paper, we extend this study by considering other quasicrystals (octagonal and Penrose tilings) described by tight-binding models. We compute their Hofstadter butterfly and study the emergence of LLs focusing on their slope and their broadening. For simplicity, we consider quasicrystals made of plaquettes with commensurate areas. They are obtained by modifying the projection direction in the standard cut-and-project (CP) method \cite{Duneau85,Kalugin85,Elser85} used in the construction of quasicrystals. For brevity, and to distinguish them from the original quasicrystals, we call them ``commensurate quasicrystals"\footnote{These commensurate quasicrystals have perfect long-range order and are no less quasicrystalline than the original ones. The adjective ``commensurate'' only refers to the fact that the ratios of their plaquette areas are rational numbers.}. In this case, the only source of irrationality in the problem is the one related to quasicrystalline order. For such commensurate quasicrystals, the Hofstadter butterfly is periodic with the magnetic flux.

 In the present work, we focus on the small-field limit where we show that LLs are generically present in quasiperiodic tight-binding models. Indeed, in the continuum limit, i.e., close to the band edges and in the zero-field limit, a quasicrystal may be described by an effective dispersion relation (which can be related to an underlying periodic crystal for the case of commensurate quasicrystals). This dispersion relation explains the existence of closed cyclotron orbits and, consequently, the emergence of LLs, their effective mass, their degeneracy and their Wilkinson-type of broadening due to quantum tunneling. For commensurate quasicrystals, the nonperiodic nature of the solid appears as a perturbation on top of the effective periodic crystal. This perturbation produces an algebraic broadening that dominates over Wilkinson's mechanism in the vanishing $f$ limit. The main result of our study is that LLs  broaden algebraically with a nonuniversal exponent.

The article is organized as follows. In Sec.~\ref{sec:models}, we introduce tight-binding models for various 2D quasicrystals and compute their Hofstadter butterfly. Then, we numerically extract the slope and broadening of LLs (Sec.~\ref{sec:hb}). Next, we propose an explanation for the surprising emergence of these LLs in quasiperiodic solids and relate their slope to an effective mass (Sec.~\ref{sec:emll}).  In Sec.~\ref{sec:bll}, we introduce models that interpolate between periodic lattices and quasicrystals and study the evolution of the LL broadening as a function of an interpolation parameter. 

%
%
\section{Hofstadter butterfly of quasicrystals}
\label{sec:models}
%
%
In this section, we introduce 2D tight-binding models on quasiperiodic tilings. More specifically, we consider tilings made of plaquettes with rational areas, such as the commensurate version of the Rauzy tiling \cite{Vidal01}, of the octagonal (or Ammann-Beenker) tiling \cite{Duneau89} and of the Penrose tiling \cite{Duneau94}. We also consider a random tiling obtained from the commensurate Rauzy tiling by phason flips (geometric disorder) \cite{Fuchs16}. We then numerically diagonalize the Hamiltonian of these tight-binding models for each flux compatible with periodic boundary conditions (PBC) in order to obtain the energy spectrum as a function of the applied perpendicular magnetic field, known as a Hofstadter butterfly (see Ref. \cite{Fuchs16} for details).

There are several reasons for considering quasicrystals made of commensurate plaquettes, rather than the original quasicrystals. The main one is that it disentangles the effects coming from incommensurate plaquettes from those that are truly due to the quasicrystalline long-range order. Note that it is easy to construct a periodic crystal made of two plaquettes with incommensurate areas. Such a system is very different from a quasicrystal, although it does have some source of irrationality. Other less important reasons to consider commensurate quasicrystals are that they possess an underlying periodic crystal (see below) and they lead to a flux-periodic Hofstadter butterfly. This last feature is not essential when discussing LLs in the vicinity of zero flux.

In practice, we consider a single unit cell of a periodic approximant\footnote{A periodic approximant is a crystal with a large unit cell that approximates a quasicrystal. It is usually part of a sequence of approximants labeled by an integer $k$, known as the order of the approximant, and containing an increasing number $N_k$ of sites in the unit cell. The quasicrystal can be seen as the $k\to \infty$ limit of the series of approximants. For a general introduction to periodic approximants of quasicrystals, see Ref.~\cite{Goldman93}.} of the quasicrystal with PBC. The zero-field Hamiltonian is
\be
H=-t\sum_{\langle i, j \rangle}|i\rangle \langle j|,
\ee
where the hopping amplitude between connected sites is taken to be $t=1$. A uniform magnetic field \mbox{$B=|\boldsymbol{\nabla}\times \boldsymbol{A}|$} is introduced as a Peierls phase on the hopping amplitudes,
\be
t_{ij}=t \to t_{ij}=t\:{\rm e}^{\im \frac{2\pi}{\phi_0} \int_{i}^j \textbf{dl}\cdot \boldsymbol{A}},
\ee
where $\boldsymbol{A}$ is the vector potential. In all tilings, we use units such that the shortest edge length is $1$, the reduced Planck constant $\hbar=1$, and the electric charge $e=2\pi$ so that the flux quantum $\phi_0=h/e=1$. We call $f=B \mathcal{A}_p$ the magnetic flux per plaquette of smallest area $\mathcal{A}_p$. Due to the PBC, the total magnetic flux across the sample \mbox{$B \mathcal{A}=N_\phi$} can only take integer values  ($\mathcal{A}$ is the total area of the system). For a convenient gauge choice adapted to this problem, we refer the interested reader to Ref.~\cite{Fuchs16}. The energy spectrum is then obtained by numerical diagonalization of the Hamiltonian as a function of the allowed fluxes $f$.

\subsection{Commensurate Rauzy tiling}
%
%
\begin{figure}[h!]
\includegraphics[width=.8\columnwidth]{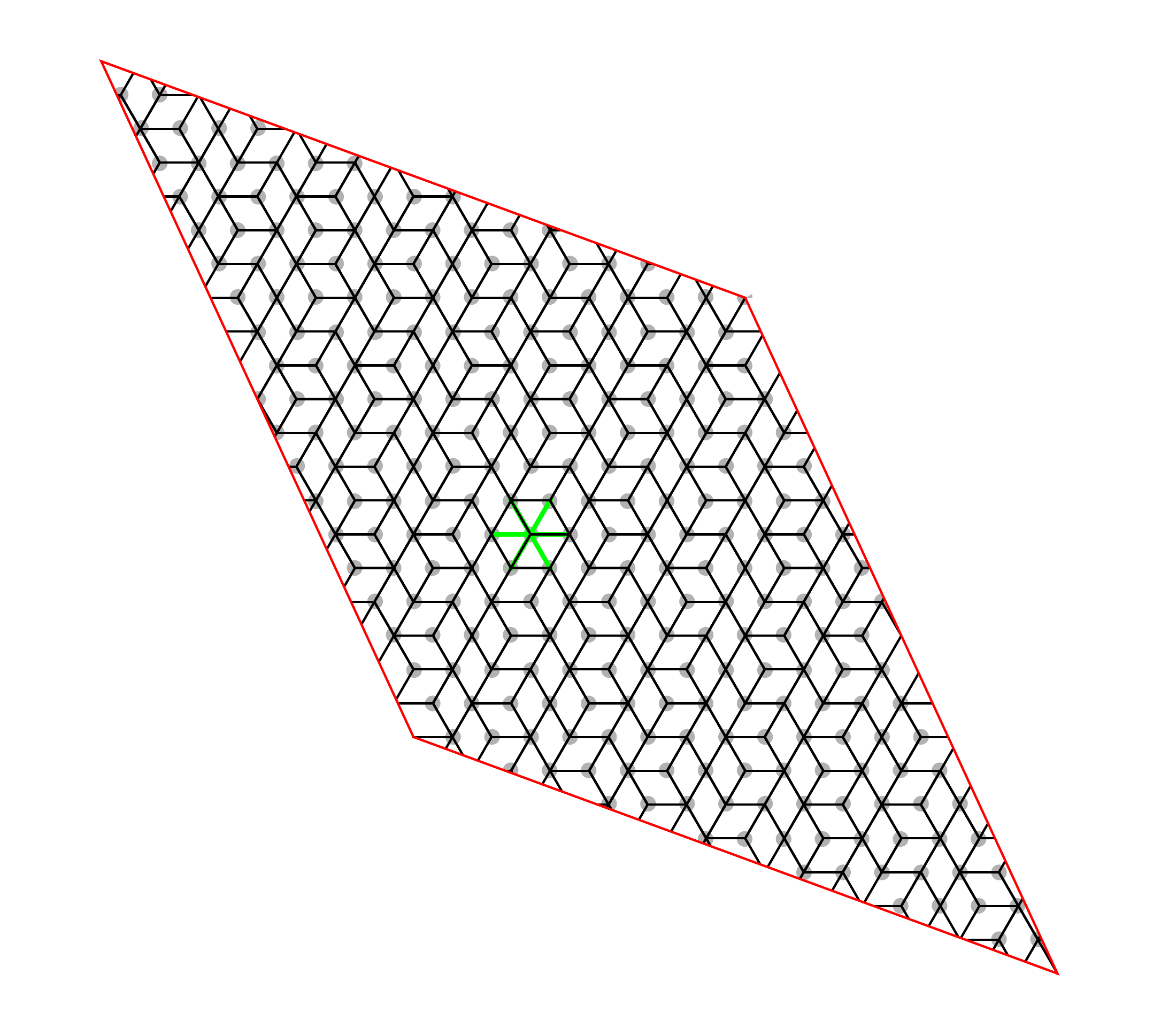}
\includegraphics[width=0.9\columnwidth]{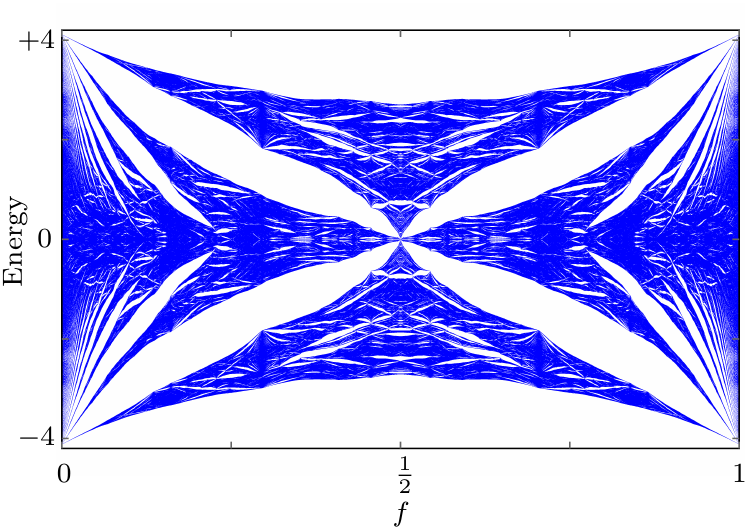}
\caption{Top: rhombic unit cell of a periodic approximant of the commensurate Rauzy tiling with 274 sites (black links). Green links and gray dots refer to the underlying periodic model discussed in Sec.~\ref{underlying} and Appendix~\ref{app:geometry}. Bottom: Hofstadter butterfly for the commensurate Rauzy tiling approximant with 5768 sites.}
\label{fig:rauzytiling}
\end{figure}
%
%
The original Rauzy tiling introduced in Ref.~\cite{Vidal01} is a codimension-one quasicrystal with three different plaquettes and the module of its Fourier transform has rank three. Its commensurate version (also known as the isometric Rauzy tiling) \cite{Vidal04} is obtained by modifying the projection direction in the construction of the original Rauzy tiling by the CP method. It contains a single rhombic plaquette of area $\sqrt{3}/2$ with three possible orientations (see Fig.~\ref{fig:rauzytiling}). Geometric details of this tiling are given in Appendix~\ref{app:geometry}. The flux per plaquette is $f=B\sqrt{3}/2$ and the allowed fluxes are $f=N_\phi/N$, where $N_\phi$ is an integer and $N$ is the number of plaquettes (it is also the number of sites). The Hofstadter butterfly has a flux periodicity $\Delta f =1$ \cite{Fuchs16} (see also Refs.~\cite{Tran15,Vidal04}). We reproduce it here for completeness (see Fig.~\ref{fig:rauzytiling}). At small $f$ and close to band edges (energy near $\pm 4.11$), LL fans separated by integer quantum Hall gaps are clearly seen. Moreover, there are much smaller gaps in the butterfly, e.g., near $f=1/2$, that are due to the quasicrystalline order and that do not feature a quantum Hall effect. For a discussion of gap labeling in the Rauzy butterfly, see Ref.~\cite{Fuchs16}.

\subsection{Phason flips and random commensurate Rauzy tiling}
%
%
\begin{figure}[t!]
\includegraphics[width=0.9\columnwidth]{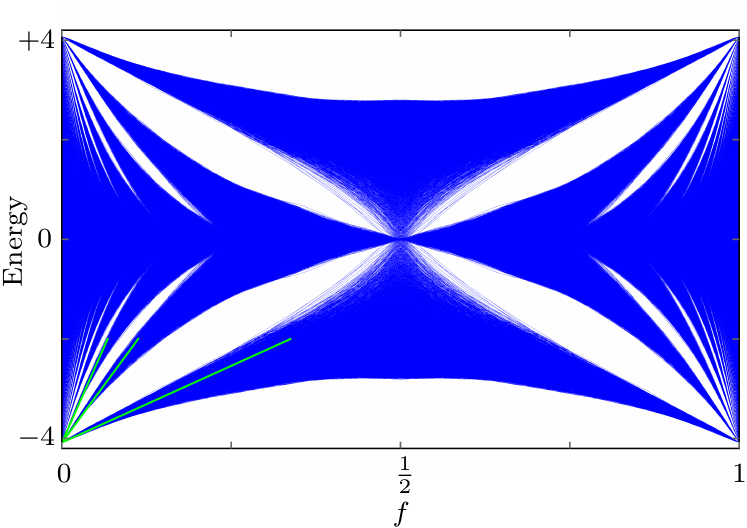}
\caption{Hofstadter butterfly for a random commensurate Rauzy tiling with 5768 sites. The three first LLs as given in Eq.~(\ref{epsnaverage}) are highlighted in green.}
\label{fig:papphason}
\end{figure}
%
%

In the Rauzy tiling discussed above, there are sites with three, four or five neighbors. Sites with three neighbors can be flipped in a process known as phason flip. Phason flips disorder the tiling. For a sample with $N$ sites, randomly flipping $N^2/2$ sites with three neighbors leads to a random tiling with the same stoichiometry (i.e., the same number of plaquettes of a given orientation) as the commensurate Rauzy tiling but without long-range order. The Hofstadter butterfly for this random tiling was obtained in Ref.~\cite{Fuchs16} and is reproduced here for completeness (see Fig.~\ref{fig:papphason}). At small $f$ and close to the band edges, broad LLs are visible (some of them are highlighted in green in Fig.~\ref{fig:papphason}). Only the main integer quantum Hall gaps survive the disorder process, but not the smaller gaps present in the nondisordered Rauzy system.

\subsection{Commensurate octagonal (Ammann-Beenker) tiling}
%
%
\begin{figure}[h!]
\includegraphics[width=0.7\columnwidth]{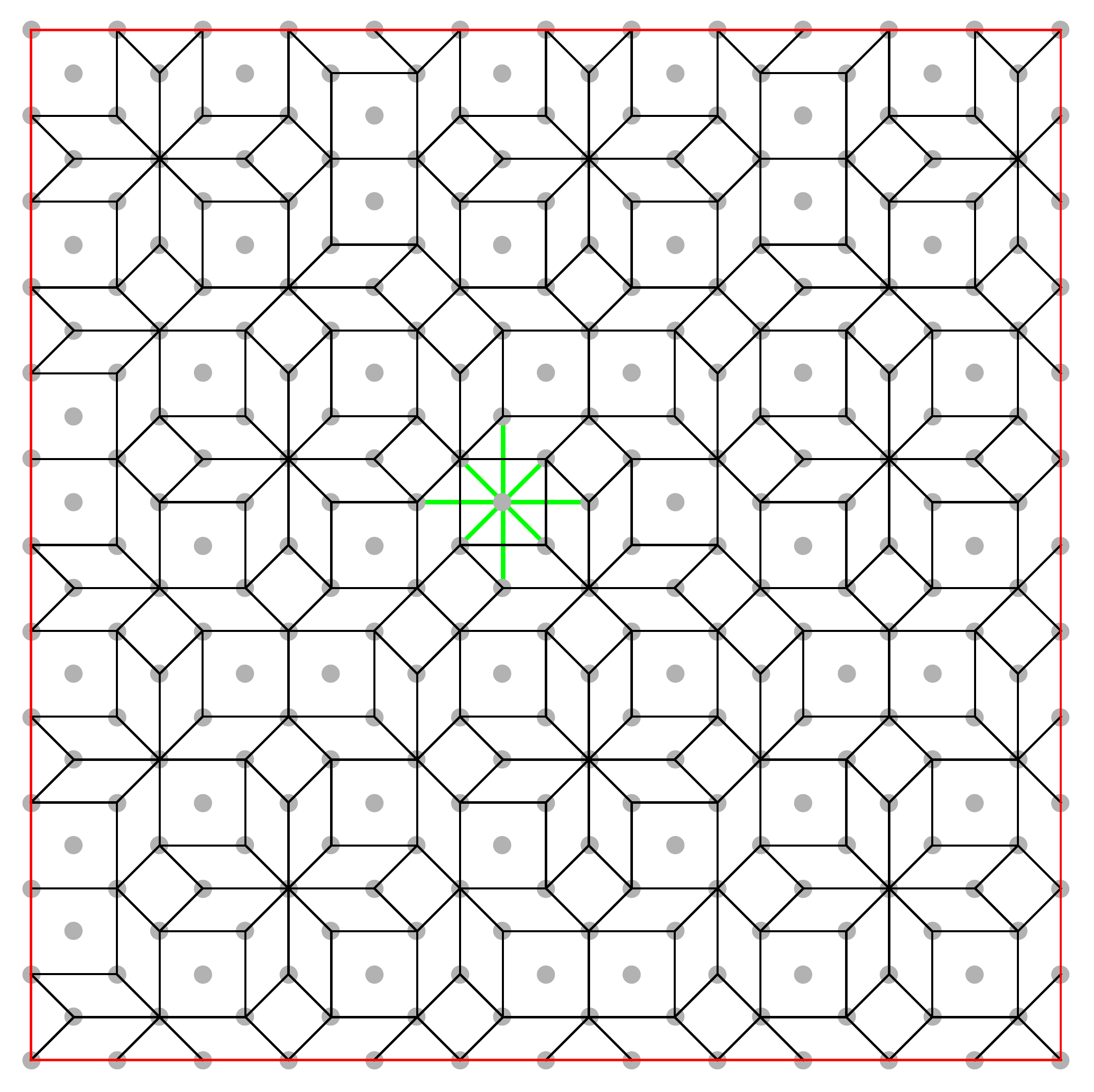}
\includegraphics[width=0.9\columnwidth]{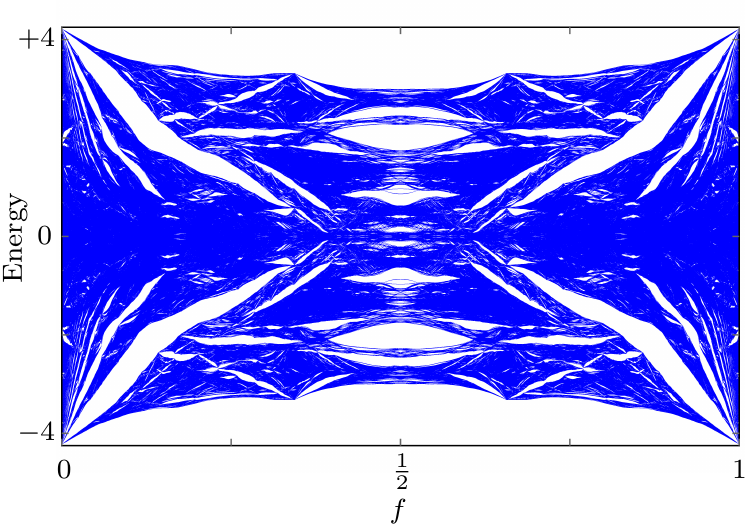}
\caption{Top: square unit cell of a periodic approximant of the commensurate octagonal tiling with 239 sites (black links). Green links and gray dots refer to the underlying periodic model discussed in Sec.~\ref{underlying} and Appendix~\ref{app:geometry}. Bottom: Hofstadter butterfly for the commensurate octagonal tiling square approximant with 8119 sites.}
\label{fig:octotiling}
\end{figure}
%
%
The octagonal (or Ammann-Beenker) \cite{Beenker82} tiling is a codimension-two tiling and the module of its Fourier transform has rank four. As for the Rauzy tiling, a commensurate version of the octagonal tiling (see Fig.~\ref{fig:octotiling}) can be obtained by modifying the projection direction in the CP method of the original tiling. Tilings considered here have one rhombic plaquette of area $1$ and two square plaquettes of area $1$ and $2$ (in units of the smallest edge length). Their construction relies on square approximants introduced in Ref.~\cite{Duneau89} (see Appendix~\ref{app:geometry}). 
The flux per smallest plaquette $f=B$ so that the allowed fluxes are $f=N_\phi/\mathcal{A}$, where $\mathcal{A}$ is the total area of the system. The corresponding butterfly is shown in Fig.~\ref{fig:octotiling} (see also \cite{Soret16}). Qualitatively, it is similar to the Rauzy butterfly, although it differs in the details. It has all the expected global symmetries: it is periodic in flux with $\Delta f =1$, has $f\to -f$ symmetry, and particle-hole symmetry, as the octagonal tiling is bipartite. It also features broad LLs near band edges at small flux and has many gaps, some of which are not related to the quantum Hall effect. Its gap labeling is left to future work.

\subsection{Commensurate Penrose tiling}
%
%
\begin{figure}[h!]
\includegraphics[width=0.8\columnwidth]{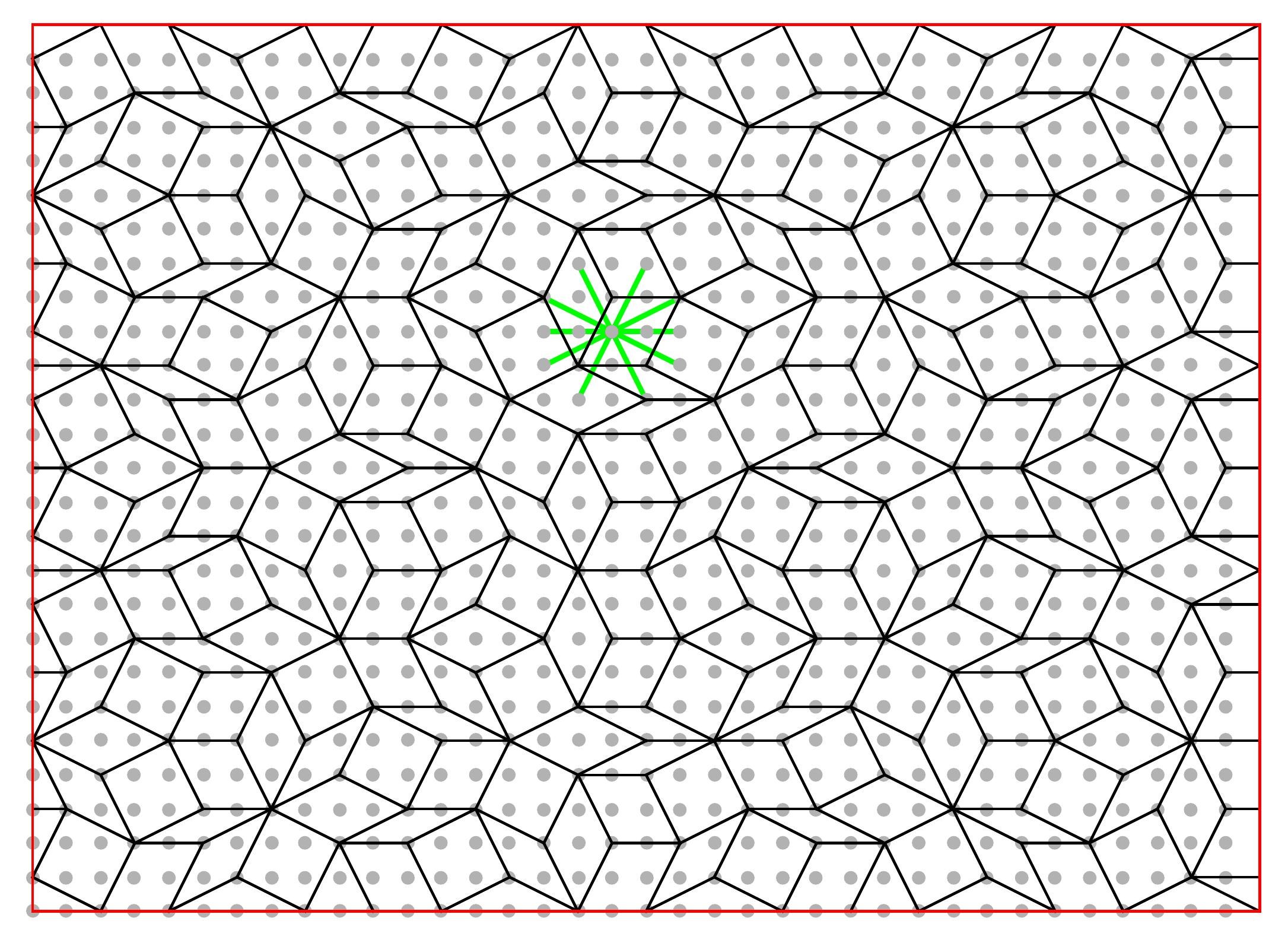}
\includegraphics[width=0.9\columnwidth]{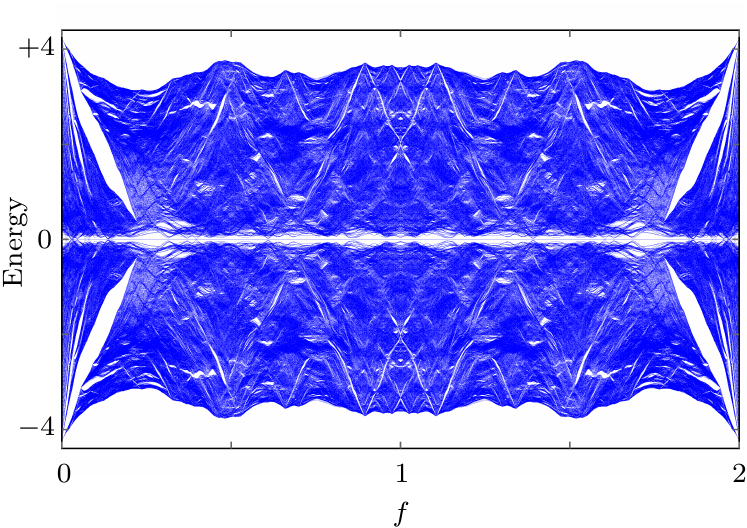}
\caption{Top: rectangular unit cell of a periodic approximant of the commensurate Penrose tiling with 246 sites (black links). Green links and gray dots refer to the underlying periodic model discussed in Sec.~\ref{underlying} and Appendix~\ref{app:geometry}. Bottom: Hofstadter butterfly of the commensurate Penrose tiling rectangular approximant with 1686 sites.}
\label{fig:penrosetiling}
\end{figure}
%
%

The Penrose rhombus tiling \cite{Penrose74} is a codimension-two tiling (even if it is frequently obtained by projecting from the $\mathbb{Z}^5$ lattice) and its Fourier transform has rank four. As for the octagonal tiling, a commensurate version of the Penrose tiling (see Fig.~\ref{fig:penrosetiling}) can be obtained by modifying the projection direction in the CP method. It has four plaquettes of area $1/2$, $3/4$, $1$, and $5/4$ (in units of the smallest edge length). Its construction is based on rectangular approximants defined in Ref.~\cite{Duneau94} (see Appendix~\ref{app:geometry} for details). 
The flux per plaquette $f=B/2$ and the allowed fluxes are $f=N_\phi/(2\mathcal{A})$. The Hofstadter butterfly is shown in Fig.~\ref{fig:penrosetiling} and has a flux periodicity of $\Delta f=2$ because $f=B/2$. The Hofstadter butterfly for the original Penrose tiling (i.e., with incommensurate plaquettes) was obtained in Ref.~\cite{Hatakeyama89}.

%
%
\section{Numerical study of Landau levels}
\label{sec:hb}
%
%
We identify LLs within the Hofstadter butterfly near band edges and at low magnetic field. A LL is a group of $N_\phi=B \mathcal{A}$ eigenenergies, which is almost degenerate and well separated from other energy levels by an energy gap. It is usual to label a LL by a non-negative integer $n$. In the following, we denote by $\epsilon_{n,1}\leq \epsilon_{n,2}\leq...\leq\epsilon_{n,N_\phi}$ the energies of the $n^\text{th}$ LL. See Figs.~\ref{fig:octotiling} and \ref{fig:lloctoslope} for the example of the commensurate octagonal tiling. At finite flux, a LL broadens and gives rise to a Landau band that features a substructure made of bands and gaps (see Fig.~\ref{fig:lloctoslope}).

\subsection{Effective mass}
\begin{figure}[t]
\includegraphics[width=\columnwidth]{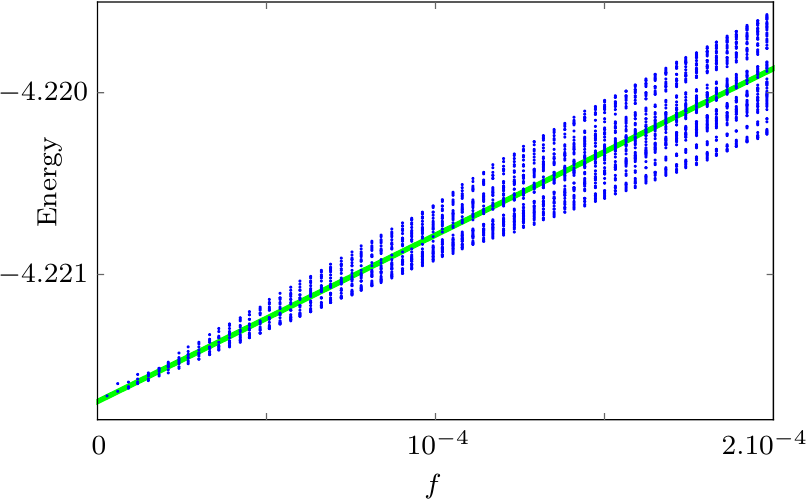}
\caption{Lowest Landau level (LLL, $n=0$, blue dots) of the commensurate octagonal tiling (approximant with 275807 sites) as a function of the flux per plaquette $f$. The green line is a linear fit to the average energy within the LLL which gives the slope $1/m$, see Table \ref{tablemass}.}
\label{fig:lloctoslope}
\end{figure}

Near a band edge with a finite zero-field density of states, one typically finds that the average energy,
\be
\langle \epsilon_n \rangle=\frac{1}{N_\phi}\sum_{j=1}^{N_\phi} \epsilon_{n,j},
\ee
of a LL behaves as
\be
\langle \epsilon_n \rangle=\text{const}+\frac{2\pi B}{m}\left(n+\frac{1}{2}\right),
\label{epsnaverage}
\ee
which allows us to extract an effective mass $m$. To this aim, we focus on the lowest Landau level (LLL) $n=0$, and check that the successive LLs are separated by an energy gap $2\pi B/m$. We consider several approximants (up to $N \sim 2 \times 10^5$ sites) and restrict the magnetic field in the range $10^{-5}\leq B \leq 10^{-3}$. The smallest non-zero flux $B$ valid on the torus is $1/\mathcal{A}$. However, the first few fluxes (say from $1/\mathcal{A}$ to $4/\mathcal{A}$) suffer from finite-size effects and do not correspond to the bulk thermodynamic energy levels \cite{Analytis05}. We therefore discard them. In the considered flux range, the LLL for the different approximants coincide, is narrow, and its average is well fitted by a linear law (see Fig.~\ref{fig:lloctoslope} for the octagonal case). Results for the inverse effective mass $1/m$; see Table \ref{tablemass}.
%
%
%
%
%
%
\begin{table}[h!]
\begin{tabular}{|c|c|c|c|c|c|}
\hline
Tiling & $1/m$& $1/m_\alpha$& $1/\widetilde{m}_\alpha$& $1/m_\rho$&$1/m_\text{T}$\\
\hline
Rauzy& $1.957(2)$&2.02&$1.97 $&$1.95(1)$&1.95735(1)\\
Octagonal  & $2.95(2)$&3.17&3&$2.88(5)$& $\to 0$\\
Penrose  & $2.40(2)$&2.51&2.37&$2.29(5)$& $\to 0$\\
Phasonized Rauzy & $1.96(1)$& 2.01 &1.97&$1.9(1)$ &$\to 0$\\
\hline
\end{tabular}
\caption{Inverse effective mass of the LLL for different tilings. Results of the first column are obtained from the slope of the LLL in the Hofstadter butterfly. Results of the  second and third columns are obtained from the effective zero-field dispersion relation. Results of the fourth column are obtained from the zero-field integrated density of states. Results of the fifth column give the Thouless inverse effective mass as obtained from the curvature of the lowest band.}
\label{tablemass}
\end{table}
%
%

\subsection{Landau level broadening}
\begin{figure}[t]
\includegraphics[width=\columnwidth]{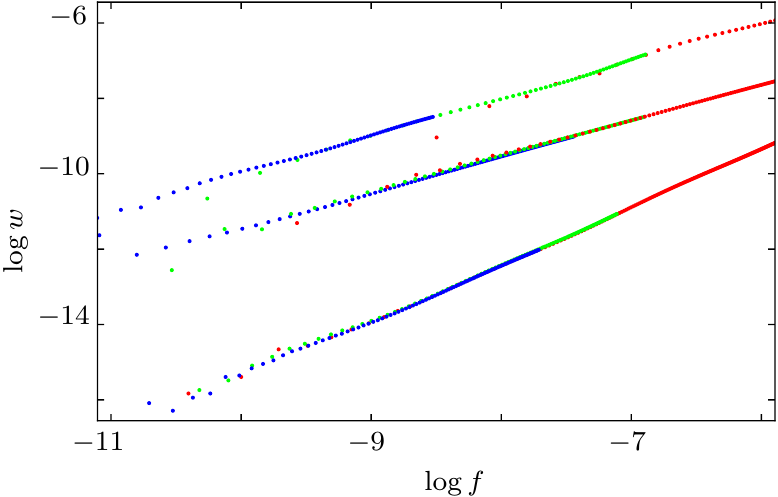}
\caption{Standard deviation $w$ of the LLL for the commensurate octagonal (top), the commensurate Penrose (middle), and the commensurate Rauzy (bottom) tilings as a function of the flux $f$ in $\log-\log$ plot. The different colors (red, green, blue) correspond to approximants with different number of sites: octagonal $(8119, 47321, 275807)$, Penrose $(30254, 79206, 207364)$, and Rauzy $(66012, 121415, 223317)$.}
\label{fig:lloctowidth}
\end{figure}

Let us now study the way a LL broadens as a function of the magnetic field. To smoothen fluctuations, we consider the width of a LL as defined by the standard deviation of all the energy states forming the Landau band considered. Here, for simplicity, we focus on the LLL for which the standard deviation reads:
\be
w=\sqrt{\langle\epsilon_{0}^2\rangle-\langle \epsilon_0 \rangle^2}.
\ee
We compute this quantity as a function of $f$ for several approximants with different number of sites and keep only the part that does not vary when changing this number of sites. We typically find a field dependence that has a power-law envelope $f^\gamma$ (see Fig.~\ref{fig:lloctowidth}). The fitted exponent $\gamma$ is given in Table \ref{tablegamma}.
%
%
%

%
%
\begin{table}[h!]
\begin{tabular}{|c|c|}
\hline
Tiling & $\gamma$\\
\hline
Rauzy & $1.5$\\
Octagonal  & $1$ \\
Penrose & $0.95$\\
Phasonized Rauzy & $0.55$\\
\hline
\end{tabular}
\caption{Exponent of the algebraic envelope of $w(f)\sim f^\gamma$. The relative error obtained by varying the fitted region is about $10\%$.}
\label{tablegamma}
\end{table}
%
%

In the case of a random tiling, one has to average over disorder realizations in order to obtain reliable results. In addition to the phasonized Rauzy tiling (see Table \ref{tablegamma}), we also studied the phasonized octagonal tiling and found $\gamma\simeq 0.5$. In both cases, the exponent is close to $\gamma=1/2$, which is the expected exponent for a non-correlated disorder in the continuum \cite{Ando74b,Wegner83}. However, phason disorder is known to be correlated since flipping one hexagon induces some constraints on its neighbors, but a complete study is beyond the scope of the present work. 
%
%
\section{Effective mass of Landau levels}
\label{sec:emll}
%
%
The main goal of this section is to understand the emergence of LLs in the Hofstadter butterfly as well as  their slope (effective mass). Since quasicrystals are not periodic, Bloch's theorem is not applicable, momentum is not conserved, and an exact dispersion relation does not exist. However, we argue below that in the continuum limit (i.e., close to band edges and in the zero-field limit), a quasicrystal is well approximated by an effective dispersion relation. In addition, when the quasicrystal is made of commensurate plaquettes, this dispersion relation can be shown to be associated to an underlying periodic crystal (see below). One may then apply the Onsager quantization condition on closed cyclotron orbits in order to obtain LLs. Below, we discuss several ways to define an effective mass accounting for the slope of LLs.

\subsection{Underlying periodic crystal and effective dispersion relation}
\label{underlying}
An effective dispersion relation $\epsilon(\mathbf{k})$ can be defined by
\be
\epsilon(\mathbf{k})=\langle \psi | H(\mathbf{k})|\psi\rangle = -\sum_{\langle i, j \rangle}{\mathrm e}^{\im \mathbf{k}\cdot \mathbf{r}_{ij}}\psi_i^* \psi_j ,
\label{eps}
\ee
where $H(\mathbf{k})={\mathrm e}^{-\im\mathbf{k}\cdot \mathbf{r}}H{\mathrm e}^{\im\mathbf{k}\cdot \mathbf{r}}$ is the Bloch Hamiltonian parametrized by a wave vector $\mathbf{k}$, $\mathbf{r}$ is the position operator, $|\psi\rangle$ is the ground state, and $\psi_j=\langle j|\psi\rangle$ the corresponding wave function. One can compute this dispersion numerically from the exact (numerical) ground state. The idea behind this definition is to have an ansatz ${\mathrm e}^{\im\mathbf{k}\cdot \mathbf{r}}|\psi\rangle$ for the long-wavelength excitations.

As we are only interested in long-wavelength properties, a cruder approximation $\widetilde{\epsilon}(\mathbf{k})$ for the dispersion relation $\epsilon(\mathbf{k})$ may be obtained by replacing the exact ground state $|\psi\rangle$ by a uniform state  $\frac{1}{\sqrt{N}}\sum_j |j\rangle$. The advantage of this approach is that it allows one to have an analytical dispersion relation,
\be
\widetilde{\epsilon}(\mathbf{k})= -\sum_{\beta} t_{\beta} \cos(\mathbf{k}\cdot \boldsymbol{\delta}_\beta),
\label{et}
\ee 
where $\boldsymbol{\delta}_\beta$ are all possible neighbor vectors labeled by $\beta$ (e.g., $\beta$ runs from $1$ to $6$ in the Rauzy tiling; see Appendix~\ref{app:geometry} for details) and $t_\beta$ is the probability to have a link along $\boldsymbol{\delta}_\beta$. The quantities $t_\beta$ play the role of effective hopping amplitudes. The average coordination number is $\bar{z}=\sum_\beta t_\beta$, which for a tiling made of rhombi is $\bar{z}=4$ so that $\widetilde{\epsilon}(0)=-4$. The above dispersion relation is periodic in reciprocal space and corresponds to a tight-binding model on a periodic lattice in real space. We refer to the latter as the ``underlying periodic crystal'' to the commensurate quasicrystal. Its construction may be seen as a geometrical mean-field-like approximation to the commensurate quasicrystal, in which each site is assumed to be identical (i.e., a Bravais lattice) and to be connected to neighbors with all the hopping terms $t_\beta$ that exist in the quasicrystal. 

For the commensurate Rauzy tiling, the effective dispersion relation (\ref{et}) is that of a triangular lattice with anisotropic nearest-neighbor hopping amplitudes (see details in Appendix~\ref{app:geometry}). The commensurate Rauzy tiling and the underlying triangular lattice have all sites in common [see Fig.~\ref{fig:rauzytiling} (top)]. In contrast, for the commensurate octagonal (Penrose) tiling, the underlying periodic crystal is a square lattice with only first- and second- (only some third- and all fourth-) nearest-neighbor hoppings. Details are given in Appendix~\ref{app:geometry}. All sites of the commensurate octagonal (Penrose) tiling belong to the underlying square lattice, but there are more sites in the square lattice, see Fig.~\ref{fig:octotiling} (top) [Fig.~\ref{fig:penrosetiling} (top)]. These two commensurate quasicrystals can therefore be seen as periodic lattices with quasiperiodic vacancies.

The energy spectrum of a quasicrystal made of rhombi is particle-hole symmetric as a result of the tight-binding model being bipartite. In contrast, the underlying periodic crystal has a dispersion relation which does not have this symmetry. Indeed, it only approximates the quasicrystal close to the band bottom.

\subsection{Effective quadratic dispersion relation}
From the effective dispersion relation $\epsilon(\mathbf{k})$ given in Eq.~(\ref{eps}) in the long-wavelength limit, one obtains
\be
\epsilon(\mathbf{k})\simeq \epsilon(0)+\frac{1}{2}\alpha_{ij} k_i k_j,
\label{eq:paraboloid}
\ee 
which defines an effective inverse mass tensor $\alpha_{ij}$. This real symmetric tensor has two positive eigenvalues $\lambda_1$ and $\lambda_2$ with $\lambda_1\geq \lambda_2$. This means that the effective dispersion relation has the shape of a paraboloid hosting closed iso-energy lines with an elliptic shape. The quantities $1/\lambda_1$ and $1/\lambda_2$ are the effective masses along the principal axes of the ellipse. The geometric mean of the eigenvalues gives an inverse effective mass $\frac{1}{m_\alpha}=\sqrt{\lambda_1 \lambda_2}$ and their ratio $\frac{\lambda_1}{\lambda_2}$ reflects the anisotropy of the tiling. A similar inverse effective mass tensor $\widetilde{\alpha}_{ij}$ can be defined from the effective dispersion relation $\widetilde{\epsilon}(\mathbf{k})$ given in Eq.~(\ref{et}). The results for the various inverse effective masses are given in Table \ref{tablemass} and those for the anisotropy in Table \ref{tableanisotropy}.
\begin{table}[h!]
\begin{tabular}{|c|c|c|c|c|}
\hline
tiling & $\lambda_1/\lambda_2$ & $\widetilde{\lambda}_1/\widetilde{\lambda}_2$\\
\hline
Rauzy & $1.48$& 1.40\\
Octagonal  & $1.0$& 1 \\
Penrose   & $1.40$ & 1.4\\
phasonized Rauzy & $1.41$& 1.40\\
\hline
\end{tabular}
\caption{Anisotropy $\lambda_1/\lambda_2$ of the inverse mass tensor $\alpha_{ij}$ obtained from the effective dispersion relation $\epsilon(\mathbf{k})$ and $\widetilde{\epsilon}(\mathbf{k})$ defined in Eqs.~(\ref{eps}) and (\ref{et}), respectively. 
}
\label{tableanisotropy}
\end{table}

 The Rauzy tiling only has a twofold rotational symmetry, which is retained by its commensurate version and which leads to an anisotropic inverse effective mass tensor (see Table \ref{tableanisotropy}). This is reflected in the underlying periodic model which has anisotropic hopping amplitudes (see Appendix~\ref{app:geometry}). Whereas the original octagonal tiling has an eightfold rotational symmetry, its commensurate version only retains a fourfold symmetry, resulting in an isotropic inverse effective mass tensor ($\lambda_1/\lambda_2=1$). In contrast, the original Penrose tiling has a tenfold symmetry, but its commensurate version only retains a twofold symmetry, resulting in an anisotropic inverse effective mass tensor ($\lambda_1/\lambda_2=1.4$). Indeed, the underlying square periodic model has very anisotropic third-nearest-neighbor hopping amplitudes (see Appendix~\ref{app:geometry}).

\subsection{Unfolding of the bands of a periodic approximant}
\label{unfolding}
A periodic approximant is a crystal with a large unit cell that approximates a quasicrystal \cite{Goldman93}. As it is a periodic system, it is described by Bloch's theorem and band theory. The corresponding energy bands are usually plotted in the first Brillouin zone (i.e., in reduced zone scheme), but can also be represented in the extended-zone scheme \cite{Kittel}. Below, we show that this unfolded band structure for an approximant suggests the existence of an effective dispersion relation for the quasicrystal, which is valid in the continuum limit and close to the ground-state energy. 

\begin{figure}[t]
\includegraphics[width=\columnwidth]{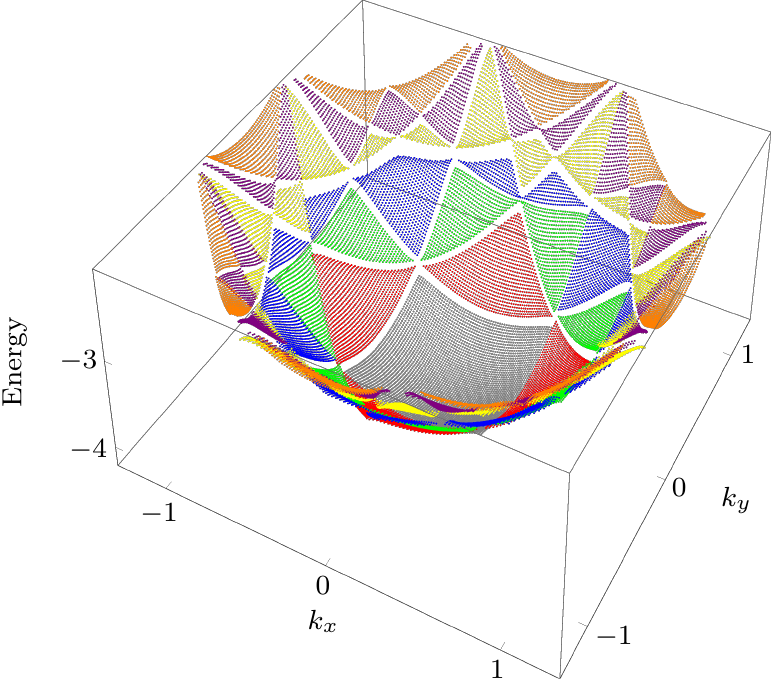}
\caption{Unfolded band structure of a periodic approximant of the commensurate octagonal tiling with $41$ sites in the unit cell. Only energies close to the ground state are shown. Colors (gray, red, green, blue, yellow, purple, orange) indicate seven bands and the corresponding Brillouin zones.}
\label{fig:unfoldedocto}
\end{figure}

In the following and for illustration, we consider the commensurate octagonal tiling and work with a periodic approximant containing $N=41$ sites in a square unit cell of area $\mathcal{A}=50$ (see Appendix~\ref{app:geometry} for details). The corresponding tight-binding model is described by a Bloch Hamiltonian $H(\mathbf{k})$, whose eigenvalues $\epsilon_b(\mathbf{k})$ define bands, where the band index $b=1,...,N$. In the reduced zone scheme, the eigenvalues $\epsilon_b(\mathbf{k})$ are plotted as a function of $\mathbf{k}$ in the first Brillouin zone, $-\pi/\sqrt{\mathcal{A}}<k_x,k_y \leq \pi/\sqrt{\mathcal{A}}$. This band structure can also be unfolded by using the higher Brillouin zones of a square Bravais lattice (extended-zone scheme). In Fig.~\ref{fig:unfoldedocto}, the unfolded band structure is plotted in the energy range  $[-4.22,-2]$. Colors indicate seven bands and the corresponding Brillouin zones. Small gaps along specific $\mathbf{k}$ directions are present at the boundaries between successive Brillouin zones\footnote{The unfolded band structure of Fig.~\ref{fig:unfoldedocto} has some similarity to Fig.~3 in Ref.~\cite{Gambaudo14}. In this reference, a very different approach is used in order to construct an effective dispersion relation for a quasicrystal. Instead of using a tight-binding model, but in the spirit of the nearly-free-electron model, the authors start from free electrons and then turn on a small potential corresponding to the sites of a Penrose tiling. Bragg peaks in the Fourier transform of the potential open gaps in the free-electron dispersion relation. These authors define modified Brillouin zones of unequal areas due to the most intense Bragg peaks, which are unlike the usual Brillouin zones that we are using.}.
%
%
%
%
%
%

\begin{figure}[t]
\includegraphics[width=\columnwidth]{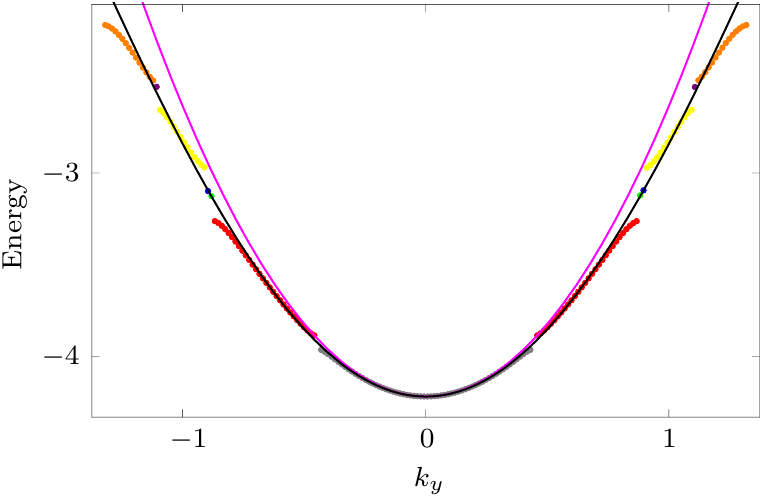}
\caption{Cut at $k_x=0$ of the unfolded band structure of Fig.~\ref{fig:unfoldedocto}. The energy is plotted as a function of $k_y$. Colors indicate seven bands and match those in Fig.~\ref{fig:unfoldedocto}. For comparison, two dispersion relations are also plotted. The effective quadratic dispersion relation $\epsilon(\mathbf{k})$ given in Eq.~(\ref{eq:paraboloid}) is plotted in magenta. The dispersion relation $\widetilde{\epsilon}(\mathbf{k})$ of the underlying square lattice [see Eq.~(\ref{etocto})] is plotted in black. It has been shifted to match the ground-state energy of the approximant.}
\label{fig:unfoldedocto_cut}
\end{figure}

Near the band bottom and close to $\mathbf{k}=0$, the unfolded band structure has an envelope which is well fitted by a paraboloid corresponding to an effective quadratic dispersion relation $\epsilon(\mathbf{k})$; see Eq.~(\ref{eq:paraboloid}). A cut of the paraboloid is plotted in Fig.~\ref{fig:unfoldedocto_cut}. There is an interesting analogy with the  ``nearly-free-electron model'' \cite{Kittel}. In the latter, one starts from free electrons described by a paraboloid and then turns on a small periodic potential that opens gaps at the boundaries between successive Brillouin zones, thereby creating a band structure in the extended-zone scheme. Here, we proceed along a reversed path: we start from a tight-binding band structure, naturally represented in the reduced-zone scheme, and unfold it in order to reveal a free-electron-like dispersion relation.

For comparison, in Fig.~\ref{fig:unfoldedocto_cut}, we also plot a $k_x=0$ cut of the dispersion relation $\widetilde{\epsilon}(\mathbf{k})$ for the underlying square tight-binding model [see Eq.~(\ref{etocto})]. This dispersion is defined in a much larger first Brillouin zone, $-\pi<k_x,k_y \leq \pi$, corresponding to a real-space microscopic periodicity of $1$, instead of the much larger periodicity $\sqrt{\mathcal{A}}$ of the periodic approximant. The dispersion relation $\widetilde{\epsilon}(\mathbf{k})$ has been shifted to match the ground-state energy of the approximant. Apart from this global shift, there are no fitting parameters. The agreement near the band bottom and in the continuum limit is very good, even beyond the paraboloid region (see Fig.~\ref{fig:unfoldedocto_cut}). 

This unfolded band structure of a periodic approximant gives a physical meaning to the effective dispersion relation of quasicrystals in the continuum limit and near the band bottom. It captures an envelope and neglects small gaps.
%
%
%
%
%
%

\subsection{Inverse effective mass of the first band, Thouless conductance and participation ratio}
The dispersion relation $\epsilon_{b=1}(\mathbf{k})$ of the first band corresponds to the lowest eigenvalues of the Bloch Hamiltonian $H(\mathbf{k})$ when $\mathbf{k}$ varies in the first Brillouin zone, $-\pi/\sqrt{\mathcal{A}}<k_x,k_y \leq \pi/\sqrt{\mathcal{A}}$. It allows one to define an effective inverse-mass tensor, 
\be
\alpha_{ij}=\left. \frac{\partial^2 \epsilon_{b=1}}{\partial k_i \partial k_j}\right|_{\mathbf{k}=0},
\ee
and a corresponding effective mass, 
\be 
m_\text{T}=1/\sqrt{\lambda_1 \lambda_2},
\ee 
where $\lambda_1$and $\lambda_2$ are the eigenvalues of $\alpha_{ij}$, (see Ref.~\cite{Fuchs16} for details). The quantity $1/m_\text{T}$ measures the sensitivity of the ground-state wave function (inside the unit cell of a periodic approximant) to the phase twist in the boundary condition as encoded by $k_x$ and $k_y$. Its physical meaning is therefore that of a dimensionless Thouless conductance for this band (hence the subscript T to the mass). The latter can be strongly influenced by the gap that opens at the edge of the first Brillouin zone and does not reveal the underlying periodic crystal but rather probes the localization of the ground-state wave function. Another useful way to characterize the localization of a wave function is the scaling exponent $\beta$ of the participation number (see Appendix~\ref{app:ipr} for details).

Results for the four tilings are given in Table \ref{tablemass}. For the Rauzy tiling, the ground-state wave function is extended ($\beta=1$) and this inverse effective mass converges to $1/m_\text{T}=1.95735(1)$ \cite{Fuchs16}. For the random tiling, the inverse effective mass goes exponentially to zero when increasing the system size, which reflects the localization by phason disorder ($\beta=0$) \cite{Fuchs16}. For the octagonal tiling, the inverse effective mass goes algebraically to zero when increasing the system size, $1/m_\text{T} \simeq 3.05/N^{0.006}$ \cite{Soret16}. This is related to the critical nature of the ground-state wave function. Indeed, the participation number scales as $N^\beta$ with $\beta\simeq 0.9877$ (see Appendix~\ref{app:ipr}). For the Penrose tiling, we find a very slow (nonmonotonic) decay of the inverse effective mass $1/m_\text{T}$ as a function of the approximant size. This is expected as the participation number exponent is very close to 1 ($\beta\simeq 0.999$; see Appendix~\ref{app:ipr}).

\subsection{Smooth integrated density of states}
Another piece of evidence for the existence of an effective dispersion relation comes from the numerically obtained zero-field integrated density of states (IDOS) for a quasicrystal, which is usually smooth near band edges. Near the ground-state energy $\ep(0)$, the IDOS per unit area can be fitted by a linear law,
\be
N_0(\ep)= \frac{m_\rho}{2\pi}[\ep-\ep(0)]\Theta[\ep-\ep(0)]
\label{IDOS}
\ee
expected for a 2D parabolic dispersion relation and characterized by a single mass parameter $m_\rho$. This effective mass is given in Table \ref{tablemass} for different tilings.

The effective dispersion relation (\ref{eq:paraboloid}) gives an IDOS of the same form as (\ref{IDOS}) with the replacement \mbox{$m_\rho \to m_\alpha=1/\sqrt{\alpha_1 \alpha_2}$}. We therefore expect $m_\alpha$ to be close to $m_\rho$.

\subsection{Semiclassical quantization of closed cyclotron orbits}
The fact that, close to a band edge, quasicrystals are qualitatively described by an effective dispersion relation with the shape of a paraboloid (see Secs. \ref{underlying} and \ref{unfolding}) is important since it explains the presence of classically \textit{closed} cyclotron orbits. The latter can then be quantized via the Onsager relation \cite{Onsager52} to give LLs as observed in the Hofstadter butterfly. According to Onsager, the zero-field IDOS (per unit area) $N_0(\ep)$ is quantized as follows
\be
N_0(\ep)=\left(n+\frac{1}{2}\right)B,
\ee
where $n$ is a non-negative integer. Note that it is not sufficient to have a finite $N_0(\ep)$ in order to obtain LLs. Indeed, open cyclotron orbits are not quantized into LLs, only closed cyclotron orbits are. 

A parabolic band bottom corresponds to an IDOS of the form given in Eq.~(\ref{IDOS}). The semiclassical quantization then gives the following LLs:
\be
\ep_n=\ep(0)+\frac{2\pi B}{m_\rho}\left(n+\frac{1}{2}\right),
\ee
which shows the relation between the slope of LLs (related to $1/m$) and the IDOS effective mass $m_\rho$. Therefore, we expect $m$ to be close to $m_\rho$. In the end, $m_\alpha\simeq m_\rho\simeq m$, as seen in Table \ref{tablemass}.

%
%
\section{Broadening of Landau levels: from periodic lattice to quasiperiodic tiling}
\label{sec:bll}
%
%

The goal of this section is to understand the observed power-law broadening $f^\gamma$ of LLs for quasicrystals. The main idea is that near band edges and in the zero-field limit, a commensurate quasicrystal can be seen as an underlying periodic crystal plus a quasiperiodic perturbation. The periodic crystal accounts for the existence of LLs, their degeneracy, and their effective mass, as discussed in the previous section. It does also account for a small Wilkinson-like broadening. However, in the vanishing $f$ limit, the latter is smaller than the LL broadening due to the quasiperiodic perturbation. The mechanism at stake is similar to that found by Rauh, except that the perturbation has a dense set of Fourier components and their interplay produces a nontrivial result: namely, it transforms an exponential into a power-law broadening.

In order to analyze this phenomenon, we introduce tight-binding models that interpolate between the quasiperiodic tiling ($\eta=1$) and the underlying periodic crystal ($\eta=0$) as a function of a parameter $\eta$ that varies some of the hopping parameters. The tight-binding Hamiltonian is taken to be
\be
H_\eta =(1-\eta) H_\text{p.}+ \eta H_\text{q.}, 
\label{interpolatingmodel}
\ee
where $H_\text{p.}$ and $H_\text{q.}$ are the Hamiltonians for the underlying periodic model and for the commensurate quasicrystal model, respectively. As soon as the dimensionless parameter $\eta\neq 0$, the system is quasiperiodic. 

\begin{figure}
\includegraphics[width=0.9\columnwidth]{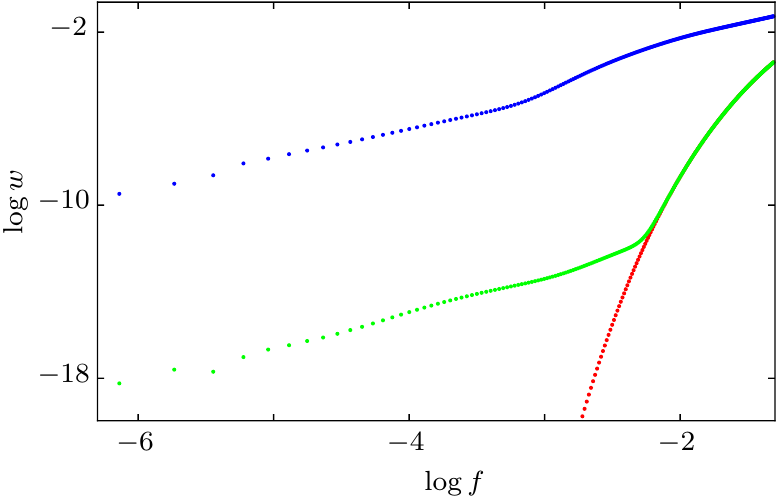}
\includegraphics[width=0.9\columnwidth]{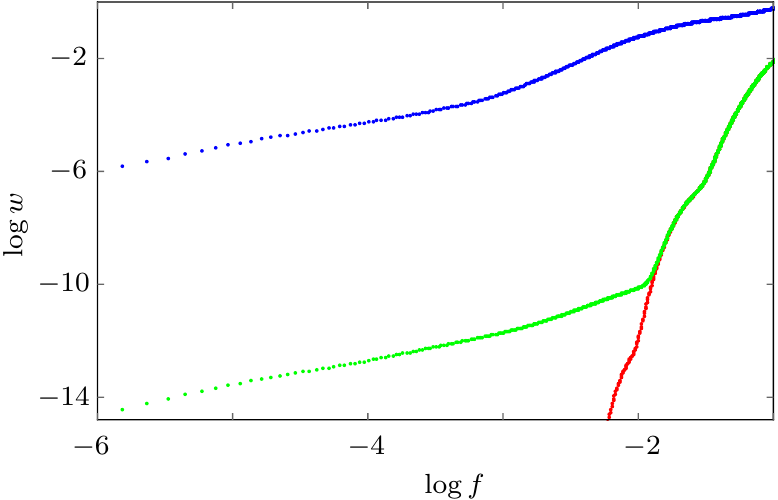}
\includegraphics[width=0.9\columnwidth]{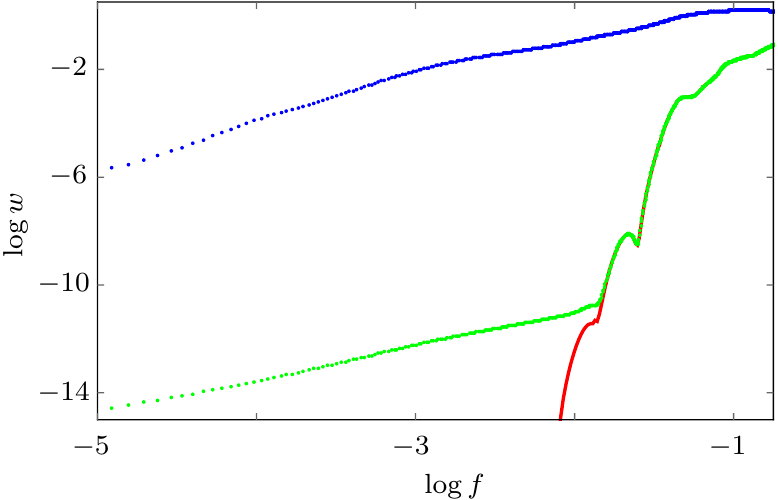}
\caption{Width $w$ of LLL as a function of the flux $f$ (log-log plot) from the Hofstadter butterfly of the periodic-quasicrystal models [see Eq.~(\ref{interpolatingmodel})] for $\eta=0$ (periodic lattice, red dots), $\eta=10^{-4}$ (green dots), and $\eta=1$ (commensurate quasicrystalline tiling, blue dots). Top: Rauzy-triangular model with 927 sites; middle: octagonal-square model with 1682 sites; bottom: Penrose-square model with 2436. Red and green points overlap when $-2\lesssim \log f$.
}
\label{fig:rauzytriangular}
\label{fig:octocarre}
\label{fig:penrosecarre}
\end{figure}

In the Rauzy case, the model interpolates between the underlying anisotropic triangular lattice and the commensurate Rauzy tiling. We find that as soon as $\eta \neq 0$, the LLL broadening in the vanishing $f$ limit is algebraic $w_\text{p}\sim \eta f^\gamma$ with an exponent $\gamma\simeq 1.5$ [see Fig.~\ref{fig:rauzytriangular} (top)]. This value of $\gamma$ appears to depend weakly on $\eta$ up to $\eta=1$. At larger flux, there is a crossover to an exponential broadening $w_\text{t}\sim \sqrt{f}\: {\mathrm e}^{-c/f}$ [see Fig.~\ref{fig:rauzytriangular} (top)], as predicted by Wilkinson for a periodic lattice. The reason for the crossover is the following. There are two broadening mechanisms in competition: one algebraic $w_\text{p}\sim \eta f^\gamma$ due to the perturbing potential and one exponential $w_\text{t}\sim \sqrt{f}\: {\mathrm e}^{-c/f}$ due to tunneling. At small flux, the algebraic broadening dominates, whereas at large flux, the exponential broadening becomes larger. The crossover occurs when the two broadenings are of similar magnitude. It appears that the total variance is approximatively given by
\be
w^2 \simeq w_\text{t}^2 + w_\text{p}^2.
\ee
%
%
%
%

In the octagonal case, the model interpolates between the underlying square lattice with first- and second-nearest neighbors and the commensurate octagonal tiling. Note that the size of the Hilbert space is given by the number of sites in the underlying periodic crystal, which is larger than that in the commensurate quasicrystal (see Appendix~\ref{app:geometry}). In the $\eta \to 1$ limit, these extra sites are weakly coupled and contribute an extensive number of near-zero-energy levels. Results for the octagonal-square model are shown in Fig.~\ref{fig:octocarre} (middle). The LLL broadening at $\eta\ll 1$ interpolates between the power-law behavior ($f^\gamma$ with $\gamma \simeq 1$) of the octagonal tiling ($\eta=1$) at small $f$ to an exponential behavior of the square lattice ($\eta=0$) when $f$ becomes larger. Even for a very small perturbation, the power-law behavior dominates in the vanishing $f$ limit and already has a similar exponent $\gamma\simeq 1$ as the $\eta=1$ limit.

In the Penrose case, the model interpolates between the underlying periodic crystal (a square lattice with some third- and all fourth-nearest neighbors) and the commensurate Penrose tiling. Results are shown in Fig.~\ref{fig:penrosecarre} (bottom) and are qualitatively similar to the octagonal-square case.

%
\section{Conclusion}
\label{sec:ccl}
%
We study the LLs that emerge in tight-binding models of quasiperiodic tilings in the limit of a weak perpendicular magnetic field. We find that their existence and their slope (i.e., inverse effective mass) can be recovered from an effective dispersion relation valid in the continuum limit. In the specific case of quasicrystals made of commensurate plaquettes, an underlying periodic crystal exists and its dispersion relation offers an interpretation of the effective dispersion relation obtained in the continuum limit. The underlying periodic crystal can be seen as a geometrical mean-field approximation to the quasiperiodic tiling. 

The nonperiodicity of the quasicrystals manifests itself in the broadening of LLs. Indeed, in the continuum, full translational symmetry ensures that LLs are infinitely narrow \cite{Landau30}. Adding a small cosine potential, a discrete translation symmetry remains which leads to an exponential broadening of LLs, which can be understood in perturbation theory  (Rauh's mechanism) \cite{Rauh75}. In a periodic crystal described by a tight-binding model, LLs are still quite narrow as they feature an exponential broadening due to tunneling, i.e., of nonperturbative origin (Wilkinson's mechanism) \cite{Wilkinson84}. We have shown that  in a 2D quasicrystal, in which translation symmetry is lost, LLs broaden with a typical width that grows roughly as a power law $f^\gamma$ with an exponent $\gamma \gtrsim 1$. In disordered systems, long-range configurational order is lost and LLs become very broad with a width that grows as the square root of the magnetic field ($\gamma=1/2$) \cite{Ando74b}. Note that for the quasicrystals that we have studied, the exponent $\gamma$ is larger than its value for disordered systems. However, we do not exclude values smaller than $1/2$.

In order to disentangle effects coming from plaquettes with incommensurate areas and true quasicrystalline long-range order, in the present paper we have chosen to focus on quasicrystals made of plaquettes with commensurate areas (which we call commensurate quasicrystals).  However, preliminary results indicate that the algebraic broadening of the LL discussed in the present work also occurs for incommensurate quasicrystals. For the three tilings studied here, the exponent $\gamma$ is found to be the same in the commensurate and in the incommensurate case (canonical construction), suggesting that this effect is not related to an underlying periodic lattice.


\medskip

\acknowledgments

We are indebted to M. Duneau for the construction of the Penrose tiling approximants. We thank G. Montambaux for many discussions on the Hofstadter butterfly, F. Pi\'echon for sharing insights on band theory and quasicrystals, and A. Soret for early work on the Hofstadter butterfly of the octagonal tiling. 
R.M. acknowledges support from the 100 talents program of Hunan Province.

%

\appendix

%
%

%
\section{Review of Rauh's approach: continuum Landau level perturbed by a cosine potential}
\label{app:rauh}
%
In this appendix, we review how a LL in the continuum is broadened by a cosine potential using first-order degenerate perturbation theory \cite{Rauh75}. The main assumption is to neglect Landau-level mixing.

Consider a periodic potential, 
\be
V(x,y)=V_0 \left[ \cos(2 \pi x/a)+\cos(2 \pi y/a) \right],
\ee 
with a single spatial frequency $1/a$. We take the following Hamiltonian:
\be
H=\frac{(\mathbf{p}+e\mathbf{A})^2}{2m} + V(x,y),
\ee
with $\mathbf{A}=B\, x\, \mathbf{u}_y$ (Landau gauge) leading to a uniform magnetic field perpendicular to the $xy$ plane.

For $V=0$,  eigenstates of $H$ in the $n^\text{th}$ LL are labeled by the wave vector $k_y$ and read
\be
\varphi_{n, k_y}(x,y)=\mathcal{N}_n \textrm{e}^{\textrm{i} k_y y} \textrm{e}^{- (x-k_y l_B^2)^2/(2l_B^2)}H_n[(x-k_y l_B^2)/l_B],
\ee
where $\mathcal{N}_n$ is a normalization factor. The corresponding eigenenergies are given by
\be
E_n=\left(n+\frac{1}{2}\right)\frac{\hbar eB}{m}.
\ee
In the following, we set the magnetic length \mbox{$l_B=\sqrt{\hbar/(eB)}=1$}. We assume a system of total area $\mathcal{A}=L_x \times L_y$ with PBC in both directions. As a consequence, $k_y=2\pi l/L_y $, where $l=1,...,N_\phi$ is an integer, with $N_\phi=\mathcal{A}/(2\pi)$ the degeneracy of a LL. 

For $0<|V_0|\ll  \frac{\hbar eB}{m}$, energy levels are obtained by diagonalizing the total Hamiltonian $H$ projected in the subspace of the $n^\text{th}$ LL. For simplicity, we restrict to the LLL ($n=0$) for which the Hermite polynomial $H_0(x)=1$ and
\be
\mathcal{N}_0=\frac{1}{\pi^{1/4}\sqrt{L_y}}, 
\ee
when $L_x\gg 1$. In this case, one has
\beqn
\langle \varphi_{0,k_y'}|V|\varphi_{0,k_y}\rangle&=&  \frac{V_0}{2}\mathrm{e}^{-\pi/(2 f)} \left[\delta_{k_y'-k_y, 2\pi/a}\right. \\
&+& \left. \delta_{k_y'-k_y, -2\pi/a}+2 \, \delta_{k_y',k_y}  \cos\left(  k_y a/f\right) \right], \nonumber
\label{eqb5}
\eeqn
where $f = a^2/(2 \pi l_B^2)=a^2/(2 \pi)$ is the dimensionless magnetic flux per unit cell of $V(x,y)$.
Because of PBC, the potential $V(x,y)$ is periodic with period $L_x$ in $x$ and $L_y$ in $y$ and its Fourier transform is discrete: it is non-zero only for discrete values of $\mathbf{q}$ such that $q_x= 2\pi n_x/L_x$ and $q_y= 2\pi n_y/L_y$ with integer $n_x$ and $n_y$. 
The energy spectrum of the $n=0$ Landau band is obtained by diagonalizing a $N_\phi \times N_\phi$ matrix, the entries of which are given by Eq.~(\ref{eqb5}).

In Eq.~(\ref{eqb5}), one recognizes the matrix elements of a 1D Harper Hamiltonian \cite{Harper55} with a frequency $1/f$ and a hopping amplitude $V_0\, \mathrm{e}^{-\pi/(2 f)}/2$. If the hopping amplitude is set to 1, this leads to an energy spectrum which is the well-known Hofstadter butterfly of the square lattice, but in a dual version, i.e., with $f$ replaced by $1/f$. This energy spectrum is bounded by -4 and 4, has a zero mean and a standard deviation of 2. The $f$-dependent hopping amplitude therefore leads to an envelope for this dual Hofstadter butterfly which is responsible for a LLL broadening,
\be
w=V_0\, {\mathrm e}^{-\pi/(2 f)},
\label{eq:wrauh}
\ee 
as found by Rauh \cite{Rauh75}. 

%
\section{Review of Wilkinson's approach: broadening of Landau levels by tunneling in the Harper equation}
\label{app:wilkinson}
%

Although the effect of a magnetic field on the energy spectrum of an electron in a tight-binding model is genuinely non-perturbative, Wilkinson found a way of studying the weak-field limit by using semiclassical techniques \cite{Wilkinson84}. Consider the 1D Harper equation describing the Hofstadter problem in the Landau gauge \cite{Harper55,Hofstadter76}. For the square lattice, it reads
\be
\epsilon \psi_m = -\psi_{m+1}-\psi_{m-1}-2\, \psi_{m} \cos(2\pi f m + k_y),
\label{eqh}
\ee
with the hopping amplitude $t=1$, the nearest-neighbor distance $a=1$, and the flux per unit cell $f=B a^2/\phi_0 = B$ with $\phi_0=1$. Wilkinson transforms it into a Schr\"odinger-like equation with an effective 1D Hamiltonian as follows. Let $X\equiv 2\pi f m + k_y$ and $P\to -i2\pi f \partial_X$ be a pair of canonically conjugate variables such that their commutator $[X,P]=\im \hbar_\text{eff}$ with $\hbar_\text{eff}=2\pi f$. Equation (\ref{eqh}) can be rewritten as
\beqn
\epsilon\: \psi(X) &=& -\psi(X+2\pi f)-\psi(X-2\pi f)-2\psi(X) \cos X ,\nonumber \\
&=& (-2\cos P -2\cos X)\psi(X),
\eeqn
thanks to $\e^{\pm \im P}\psi(X)=\psi(X\pm2\pi f)$ with $\psi(X)=\psi_m$. We therefore obtain the following Wilkinson Hamiltonian: 
\be
H(X,P)=-2\cos P -2\cos X .
\ee
It is a 1D Hamiltonian in the continuum and with an unusual kinetic energy $-2\cos P$ and a potential energy $-2\cos X$. The semiclassical limit $\hbar_\text{eff}\to 0$ corresponds to the weak-field limit. Note that the zero-field limit corresponds to a classical model written in $X,P$ phase space with Poisson bracket $\{X,P\}=1$. 

Starting from the classical limit, the first level of approximation consists in using Bohr-Sommerfeld quantization to study the Wilkinson Hamiltonian. This leads to isolated LLs near the minima of each potential well of $-2\cos X$. Indeed, near a minimum, \mbox{$H\simeq -4 + X^2 + P^2$}, the surface enclosed by an iso-energy curve $H=\epsilon$ is $S(\epsilon)=\oint_{H=\epsilon} \text{d}X\,  P= \pi(\epsilon+4)$. The Bohr-Sommerfeld relation $S(\epsilon)=2\pi \hbar_\text{eff} (n+1/2)$ gives 
\mbox{$\epsilon_n=-4+4\pi f (n+1/2)$}. At fixed index $n$, these LLs are degenerate (there is one such state per well in phase space). At this level of approximation, there is no broadening. 

A finite width of the LLs appears at the next order of approximation, when one introduces quantum tunneling, also known as magnetic breakdown, between neighboring wells of the potential $-2\cos X$. The WKB approximation gives the following tunneling amplitude \cite{LL3}:
\be
t_n = \frac{\hbar_\text{eff}}{|S'(\epsilon_n)|} \e^{-\frac{1}{\hbar_\text{eff}}\int_{X_n}^{2\pi-X_n} dX \text{Im} P(X)},
\ee
where $H(X_n,P)=\epsilon_n$, which gives $X_n\simeq \sqrt{4\pi f (n+1/2)}$ and $\text{Im } P(X)=\text{argcosh} (-\frac{\epsilon_n}{2}-\cos X)$. For the LLL ($n=0$), computing the integral numerically, we find that
\be
t_0 \simeq 7.5 \sqrt{f}\:\e^{-4\beta(2)/(\pi f)},
\ee
where $\beta(2)\simeq 0.915965$ is the Catalan constant. 

The physical picture is that LLs localized near the minimum of each well in phase space form a 2D square lattice and are coupled between neighboring wells by a hopping amplitude $t_0$. The bandwidth of such a square tight-binding model in phase space is $8 t_0$ and the corresponding standard deviation is $w=2t_0$.

The next level of approximation, which is not needed for our purpose but which we mention for completeness, is that the hopping amplitudes between LLs are actually complex numbers, i.e., they carry phases. This means that the square tight-binding model in phase space is actually in an effective magnetic field. The corresponding phase that is accumulated around a square plaquette in phase space is found to be proportional to $1/f$. This results in the emergence of a dual Hofstadter butterfly within the broadened Landau band. This process then continues and leads to the self-similar structure of the Hofstadter butterfly, as understood by Wilkinson \cite{Wilkinson84}.

The above result should be compared to that found by Rauh \cite{Rauh75} starting from continuum LLs (see Appendix~\ref{app:rauh}). Rauh finds that the width of the LLL is given by Eq.~(\ref{eq:wrauh}). There are two main differences. First, the numerical coefficient in the exponential is different: $\pi/2$ versus $4\beta(2)/\pi$. Second, the preexponential power-law behavior differs: $f^0$ versus $f^{1/2}$. Numerical calculations on the Hofstadter butterfly unambiguously show that Wilkinson's result is correct. The difficulty with Rauh's calculation is that it is performed in the continuum and therefore assumes that the perturbing potential varies on a large scale compared to the lattice spacing.

%
\section{Geometric details for the periodic approximants of commensurate quasicrystals and their underlying periodic lattice}
\label{app:geometry}
%

In this appendix, we give geometrical details of the construction of periodic approximants to the commensurate quasicrystals studied. They are obtained by the standard CP method \cite{Duneau85,Kalugin85,Elser85}. The selection (or cut) procedure is the same for the commensurate quasicrystals as for the original quasicrystals. However, the projection direction is modified such that the ratios of the plaquette areas are rational numbers. 

We also give the tight-binding model of the underlying periodic crystals and the corresponding dispersion relation. 

In the following and for all tilings, we take the shortest edge length to have unit length.

\subsection{Commensurate Rauzy tiling and underlying triangular lattice}
The original Rauzy tiling and its rhombic approximants obtained by a CP method from $3$D to $2$D are described in Ref.~\cite{Vidal01}. Its commensurate version (also known as the isometric Rauzy tiling) is obtained by modifying the projection direction \cite{Vidal04}. Edges in the $\mathbb{Z}^3$ lattice --- as represented by three canonical basis vectors --- are transformed into edges in the tiling plane $\mathbb{R}^2$ as follows:
\beqn
(1,0,0)&\to&  (1,0)= \boldsymbol{\delta}_1=-\boldsymbol{\delta}_4, \nonumber \\
(0,1,0)&\to&(-1/2, \sqrt{3}/2)=\boldsymbol{\delta}_3=-\boldsymbol{\delta}_6,\nonumber \\
(0,0,1)&\to&  (-1/2,-\sqrt{3}/2)=\boldsymbol{\delta}_5=-\boldsymbol{\delta}_2.
\label{rauzynn}
\eeqn
Selected points in the cubic lattice have integer coordinates $(p,q,r)$ which are transformed into coordinates $(p-q/2-r/2,q\sqrt{3}/2-r\sqrt{3}/2)$ in the tiling plane thanks to the above rule. Note that in the commensurate quasicrystal tiling, all edges have the same length. For the $k^\text{th}$ approximant, the number of sites is \mbox{$N=R_{k+1}$}, where $R_k$ is a Rauzy number defined by 
\be 
R_k=R_{k-1}+R_{k-2}+R_{k-3},
\ee 
with $R_2=2$, $R_1=1$, $R_0=1$. The ratio $R_{k}/R_{k-1}$ tends to the Tribonacci constant $\rho \simeq1.83929$ when $k\to \infty$. 

In the commensurate Rauzy tiling, there is a single rhombic plaquette of area $\mathcal{A}_p=\sqrt{3}/2$ with three possible orientations (see Fig.~\ref{fig:rauzytiling}). The total area is simply \mbox{$\mathcal{A}=R_{k+1} \sqrt{3}/2$}.

The underlying triangular lattice has basis vectors $\mathbf{a}_1=(1/2,\sqrt{3}/2)$ and $\mathbf{a}_2=(-1/2,\sqrt{3}/2)$ and  nearest-neighbor vectors $\boldsymbol{\delta}_\beta$ with $\beta=1,...,6$ as given in Eq.~(\ref{rauzynn}). These six hopping directions are indicated as green links in Fig.~\ref{fig:rauzytiling} (top). The anisotropic hopping amplitudes are $t_1=t_4=1-\rho^{-1}$, $t_2=t_5=1-\rho^{-2}$, $t_3=t_6=1-\rho^{-3}$. The dispersion relation of the underlying triangular tight-binding model given in Eq.~(\ref{et}) becomes:
\beqn
\widetilde{\epsilon}(\mathbf{k})&=&-(1-\rho^{-1})\cos k_x-(1-\rho^{-2})\cos \frac{k_x+\sqrt{3}k_y}{\sqrt{2}}\nonumber \\
&-&(1-\rho^{-3})\cos \frac{k_x-\sqrt{3}k_y}{\sqrt{2}}.
\eeqn

\subsection{Commensurate octagonal tiling and underlying square lattice}
The original octagonal tiling is described in Ref.~\cite{Beenker82} and its square approximants obtained by a CP method from $4$D to $2$D are defined in Ref.~\cite{Duneau89}. Here, we consider the same selection procedure in the $\mathbb{Z}^4$ lattice and a modified projection direction that gives three plaquettes of commensurate area: a rhombus of area $1$ and two squares of area $1$ and $2$ (see Fig.~\ref{fig:octotiling}). Edges in the $\mathbb{Z}^4$ lattice are transformed into edges in the tiling plane $\mathbb{R}^2$ as follows:
\beqn
(1,0,0,0)&\to& (\sqrt{2},0)=\boldsymbol{\delta}_1=-\boldsymbol{\delta}_5, \nonumber \\
(0,1,0,0)&\to&(0, \sqrt{2})=\boldsymbol{\delta}_3=-\boldsymbol{\delta}_7,\nonumber \\
(0,0,1,0)&\to& (1/\sqrt{2}, 1/\sqrt{2})=\boldsymbol{\delta}_2=-\boldsymbol{\delta}_6,\nonumber \\
(0,0,0,1)&\to& (-1/\sqrt{2}, 1/\sqrt{2})=\boldsymbol{\delta}_4=-\boldsymbol{\delta}_8 .
\label{octonn}
\eeqn
Note that, in this deformed tiling, there are both links of length $1$ ($\delta_\beta$ with even $\beta$) and of length $\sqrt{2}$ ($\delta_\beta$ with odd $\beta$). For the $k^\text{th}$ square approximant of the octagonal tiling, the number of sites is $N=O_{2k+1}+O_{2k}$ and the vectors of the unit cell are $\mathbf{A}_1=(\sqrt{2}O_{k+1},0)$ and $\mathbf{A}_2=(0,\sqrt{2}O_{k+1})$, where $O_k$ are Pell (or octonacci) numbers obeying 
\be
O_{k}=2O_{k-1}+O_{k-2},
\ee 
with $O_2=2$ and $O_1=1$. The ratio $O_{k}/O_{k-1}$ tends to the silver mean $1+\sqrt{2}$ when $k\to \infty$. The total area is \mbox{$\mathcal{A}=2\, O_{k+1}^2$} and the average area per site \mbox{$\mathcal{A}/N\to (1+\sqrt{2})/2$} in the thermodynamic limit.

The underlying periodic structure is a square lattice with basis vectors $\mathbf{a}_1=(1,1)/\sqrt{2}$ and $\mathbf{a}_2=(-1,1)/\sqrt{2}$ and  hopping along $\boldsymbol{\delta}_\beta$ with $\beta=1,...,8$ as given in Eq.~(\ref{octonn}). For even $\beta$, it corresponds to nearest neighbors, while for odd $\beta$, it corresponds to next-nearest neighbors. These eight hopping directions are indicated as green links in Fig.~\ref{fig:octotiling} (top). All hopping amplitudes $t_\beta=1/2$ are equal due to the eightfold symmetry of the original octagonal tiling. The dispersion relation of the underlying square tight-binding model [see Eq.~(\ref{et})] is:
\beqn
\widetilde{\epsilon}(\mathbf{k})&=&-\frac{1}{2}\left[\cos \frac{k_x+k_y}{\sqrt{2}}+\cos\frac{k_x-k_y}{\sqrt{2}}\right. \nonumber \\
&+&\left.\cos(k_x \sqrt{2})+\cos(k_y \sqrt{2})\right] .
\label{etocto}
\eeqn
All sites of the commensurate octagonal tiling belong to the square lattice, but there are $(1+\sqrt{2})/2$ times more sites in the square lattice (see Fig.~\ref{fig:octotiling}).  This number is the ratio between the average area per site in the octagonal tiling, $\mathcal{A}/N$, and that in the square lattice, $|\mathbf{a}_1| |\mathbf{a}_2|=1$.

\subsection{Commensurate Penrose tiling and underlying square lattice}
The original Penrose tiling is defined in Ref.~\cite{Penrose74} and its rectangular (Taylor) approximants obtained by the CP method from $5$D to $2$D are described in Ref.~\cite{Duneau94}. Here, we consider the same selection procedure in the $\mathbb{Z}^5$ lattice and modify the projection direction such as to obtain four plaquettes of commensurate areas: three rhombi of area $1/2$, $3/4$, and $1$ and a square of area $5/4$; see Fig.~\ref{fig:penrosetiling}. Edges in the $\mathbb{Z}^5$ lattice are transformed into edges in the tiling plane $\mathbb{R}^2$ as follows:

\beqn
(1,0,0,0,0)&\to& (1,0)=\boldsymbol{\delta}_1=-\boldsymbol{\delta}_6, \nonumber \\
(0,1,0,0,0)&\to&(1/2, 1)=\boldsymbol{\delta}_3=-\boldsymbol{\delta}_8,\nonumber \\
(0,0,1,0,0)&\to& (-1,1/2)=\boldsymbol{\delta}_5=-\boldsymbol{\delta}_{10},\nonumber \\
(0,0,0,1,0)&\to&(-1,-1/2)=\boldsymbol{\delta}_7=-\boldsymbol{\delta}_2,\nonumber \\
(0,0,0,0,1)&\to&  (1/2,-1)=\boldsymbol{\delta}_9=-\boldsymbol{\delta}_4 .
\label{penrosenn}
\eeqn
Note that there are both links of length $1$ ($\delta_\beta$ with $\beta=1,6$) and of length $\sqrt{5}/2$ ($\beta=2,3,4,5,7,8,9,10$). For the $k^\text{th}$ approximant of the Penrose tiling, the number of sites is $N=f_{k+3}$, the vectors of the unit cell are $\mathbf{A}_1=(\mathcal{F}_{k+3},0)$ and $\mathbf{A}_2=(0,F_{k+3})$, and the total area is $\mathcal{A}= \mathcal{F}_{k+3} F_{k+3}$, where $F_k$, $f_k$, and $\mathcal{F}_k$ obey the Fibonacci recursion relation
\be
F_k=F_{k-1}+F_{k-2},
\ee 
but with different initial conditions ($f_2=22$, $f_1=14$; $F_2=2$, $F_1=1$; and $\mathcal{F}_2=3$, $\mathcal{F}_1=1$). The $F_k$ are the canonical Fibonacci numbers and the ratio $F_{k}/F_{k-1}$ tends to the golden mean $(1+\sqrt{5})/2$ when $k\to \infty$. The average area per site is $\mathcal{A}/N\to 1/2+1/\sqrt{5}$ in the thermodynamic limit. 

The underlying periodic structure is a square lattice with basis vectors $\mathbf{a}_1=(1/2,0)$ and $\mathbf{a}_2=(0,1/2)$ and hopping along $\boldsymbol{\delta}_\beta$ (with $\beta=1,...,10$) as given in Eq.~(\ref{penrosenn}). The ten hopping directions are indicated as green links in Fig.~\ref{fig:penrosetiling} (top). They correspond to only some of the third neighbors ($\beta=1,6$) and all of the fourth neighbors ($\beta=2,3,4,5,7,8,9,10$). The hopping amplitudes are all equal $t_\beta =2/5$ due to the tenfold symmetry of the original Penrose tiling. The dispersion relation of the underlying square tight-binding model [see Eq.~(\ref{et})] is:
\beqn
\widetilde{\epsilon}(\mathbf{k})&=&-\frac{2}{5}\left[\cos k_x+\cos\frac{2k_x+k_y}{2}+\cos\frac{k_x+2k_y}{2}\right. \nonumber \\
&+&\left.\cos\frac{k_x-2k_y}{2}+\cos\frac{2k_x-k_y}{2}\right] .
\eeqn
All sites of the commensurate Penrose tiling belong to the square lattice, but there are $2+4/\sqrt{5}$ times more sites in the square lattice (see Fig.~\ref{fig:penrosetiling}). This number is the ratio between the average area per site in the Penrose tiling, $\mathcal{A}/N$, and that in the square lattice $|\mathbf{a}_1| |\mathbf{a}_2|=1/4$.

%
\section{Scaling of the participation number for the zero-field groundstate of 2D quasicrystals}
\label{app:ipr}
%
 A useful way to characterize the localization of a state $|\psi\rangle=\sum_j \psi_j |j\rangle$ (which we take to be normalized) is through the scaling of the participation number $P$, i.e., the number of sites supporting the wave function and defined by \cite{Grimm03}
\be
P = \frac{1}{\sum_{j=1}^N |\psi_j|^4},
\ee
where $N$ is the total number of sites. When $N\to \infty$, $P$ behaves as $N^\beta$, where $\beta=1$ for an extended wave function, $\beta=0$ for a localized wave function, and $0<\beta<1$ for a critical wave function. For example, in 2D, a wave function with a long-distance power-law decay $1/|\mathbf{r}-\mathbf{r}_0|^\alpha$ has $\beta=0$ if $\alpha\geq 1$, $\beta=1$ if $\alpha\leq 1/2$, and $0<\beta=2(1-\alpha)<1$ if $1/2<\alpha<1$. 

For the Rauzy tiling, the zero-field ground-state wave function is numerically found to have a scaling exponent that converges to $\beta=1$ when increasing the approximant size. More precisely, we find $\beta=0.99999995(1)$ by studying approximants with up to $N=223\, 317$ sites. This is the behavior of an extended wave function. 

For the Penrose tiling, we find that $\beta =0.998(2)$, by studying approximants with up to $N=1\, 421\, 294$ sites. For the octagonal tiling, we find that $\beta=0.98774(2)$, by studying approximants with up to $N=1\, 607\, 521$ sites. As $\beta<1$, these examples correspond to critical wave functions.

These results agree with an analytical expression for the scaling exponent $\beta$ \cite{Mace17} that can be derived from the Sutherland-Kalugin-Katz expression for the ground-state wave function of certain 2D quasicrystals \cite{Sutherland86a,Kalugin14}. Sutherland gives an exact expression for the scaling exponent in the case of the Penrose tiling as \cite{Sutherland87}
\be
\beta=\alpha(2\kappa)-\alpha(4\kappa)/2,
\label{betasutherland}
\ee
with 
\be
\alpha(\kappa)=\frac{\ln\left[\cosh \kappa +\frac{5}{2}+\sqrt{(\cosh \kappa +\frac{7}{2})(\cosh \kappa +\frac{3}{2})}\right]-\kappa}{\ln^2 \tau},
\label{alphasutherland}
\ee
where the golden mean $\tau=(1+\sqrt{5})/2$ is the inflation factor of the tiling and $\kappa$ is a scaling factor for the wave function [note that in Eq.~(42) of Ref.~\cite{Sutherland87} giving the function $\alpha(\kappa)$, there is a misplaced parenthesis]. The scaling factor $\kappa$ is related to $\lambda=\mathrm{e}^{2\kappa}$ defined in Ref.~\cite{Kalugin14}, where it was numerically found that $\lambda=1.07500(1)$ for the ground-state wave function of the Penrose tiling. When plugged in (\ref{betasutherland}) and (\ref{alphasutherland}), this gives $\beta=0.9991897(2)$, in agreement with the above numerical result.

The scaling exponent $\beta$ is called $d_2(\psi)$ in Ref.~\cite{Mace17} and given by
\be
\beta = \ln \left(\frac{\omega(2\kappa)^2}{\omega(4\kappa)} \right)/\ln \omega(0),
\label{betamace}
\ee
in terms of a function $\omega(\kappa)$ that depends on the tiling. 

For the Penrose tiling, one has
\be
\omega(\kappa)=\frac{b(\kappa)+\sqrt{b(\kappa)^2-4\mathrm{e}^{2\kappa}}}{2},
\ee
with $b(\kappa)=\mathrm{e}^{2\kappa}+5\mathrm{e}^\kappa+1$ \cite{Mace17}, which exactly matches Sutherland's result \cite{Sutherland87}.

The same expression (\ref{betamace}) holds for the octagonal tiling but with
\be
\omega(\kappa)=\frac{a(\kappa)+\sqrt{a(\kappa)^2-\mathrm{e}^{2\kappa}}}{\mathrm{e}^\kappa},
\ee
where $a(\kappa)=4\mathrm{e}^{2\kappa}+9\mathrm{e}^\kappa+4$ \cite{Mace17}. 
Using the scaling factor $\lambda=\e^{2\kappa}=1.358076(2)$ \cite{Kalugin14}, one gets $\beta=0.9877533(1)$, in agreement with the above numerical result.


\begin{thebibliography}{30}%
\makeatletter
\providecommand \@ifxundefined [1]{%
 \@ifx{#1\undefined}
}%
\providecommand \@ifnum [1]{%
 \ifnum #1\expandafter \@firstoftwo
 \else \expandafter \@secondoftwo
 \fi
}%
\providecommand \@ifx [1]{%
 \ifx #1\expandafter \@firstoftwo
 \else \expandafter \@secondoftwo
 \fi
}%
\providecommand \natexlab [1]{#1}%
\providecommand \enquote  [1]{``#1''}%
\providecommand \bibnamefont  [1]{#1}%
\providecommand \bibfnamefont [1]{#1}%
\providecommand \citenamefont [1]{#1}%
\providecommand \href@noop [0]{\@secondoftwo}%
\providecommand \href [0]{\begingroup \@sanitize@url \@href}%
\providecommand \@href[1]{\@@startlink{#1}\@@href}%
\providecommand \@@href[1]{\endgroup#1\@@endlink}%
\providecommand \@sanitize@url [0]{\catcode `\\12\catcode `\$12\catcode
  `\&12\catcode `\#12\catcode `\^12\catcode `\_12\catcode `\%12\relax}%
\providecommand \@@startlink[1]{}%
\providecommand \@@endlink[0]{}%
\providecommand \url  [0]{\begingroup\@sanitize@url \@url }%
\providecommand \@url [1]{\endgroup\@href {#1}{\urlprefix }}%
\providecommand \urlprefix  [0]{URL }%
\providecommand \Eprint [0]{\href }%
\providecommand \doibase [0]{http://dx.doi.org/}%
\providecommand \selectlanguage [0]{\@gobble}%
\providecommand \bibinfo  [0]{\@secondoftwo}%
\providecommand \bibfield  [0]{\@secondoftwo}%
\providecommand \translation [1]{[#1]}%
\providecommand \BibitemOpen [0]{}%
\providecommand \bibitemStop [0]{}%
\providecommand \bibitemNoStop [0]{.\EOS\space}%
\providecommand \EOS [0]{\spacefactor3000\relax}%
\providecommand \BibitemShut  [1]{\csname bibitem#1\endcsname}%
\let\auto@bib@innerbib\@empty
\bibitem [{\citenamefont {Onsager}(1952)}]{Onsager52}%
  \BibitemOpen
  \bibfield  {author} {\bibinfo {author} {\bibfnamefont {L.}~\bibnamefont
  {Onsager}},\ }\bibfield  {title} {\enquote {\bibinfo {title} {{Interpretation
  of the de Haas--van Alphen effect}},}\ }\href {\doibase
  10.1080/14786440908521019} {\bibfield  {journal} {\bibinfo  {journal}
  {Philos. Mag.}\ }\textbf {\bibinfo {volume} {43}},\ \bibinfo {pages} {1006}
  (\bibinfo {year} {1952})}\BibitemShut {NoStop}%
\bibitem [{\citenamefont {Landau}(1930)}]{Landau30}%
  \BibitemOpen
  \bibfield  {author} {\bibinfo {author} {\bibfnamefont {L.}~\bibnamefont
  {Landau}},\ }\bibfield  {title} {\enquote {\bibinfo {title} {{Diamagnetism of
  metals}},}\ }\href {\doibase 10.1007/BF01397213} {\bibfield  {journal}
  {\bibinfo  {journal} {Z. Phys.}\ }\textbf {\bibinfo {volume} {64}},\ \bibinfo
  {pages} {629} (\bibinfo {year} {1930})}\BibitemShut {NoStop}%
\bibitem [{\citenamefont {Ando}(1974)}]{Ando74b}%
  \BibitemOpen
  \bibfield  {author} {\bibinfo {author} {\bibfnamefont {T.}~\bibnamefont
  {Ando}},\ }\bibfield  {title} {\enquote {\bibinfo {title} {{Theory of Quantum
  Transport in a Two-Dimensional Electron System under Magnetic Fields. III.
  Many-Site Approximation}},}\ }\href {\doibase 10.1143/JPSJ.37.622} {\bibfield
   {journal} {\bibinfo  {journal} {J. Phys. Soc. Japan}\ }\textbf {\bibinfo
  {volume} {37}},\ \bibinfo {pages} {622} (\bibinfo {year} {1974})}\BibitemShut
  {NoStop}%
\bibitem [{\citenamefont {Rauh}(1975)}]{Rauh75}%
  \BibitemOpen
  \bibfield  {author} {\bibinfo {author} {\bibfnamefont {A.}~\bibnamefont
  {Rauh}},\ }\bibfield  {title} {\enquote {\bibinfo {title} {{On the Broadening
  of Landau Levels in crystals}},}\ }\href {\doibase 10.1002/pssb.2220690137}
  {\bibfield  {journal} {\bibinfo  {journal} {Phys. Stat. Sol. (b)}\ }\textbf
  {\bibinfo {volume} {69}},\ \bibinfo {pages} {K9} (\bibinfo {year}
  {1975})}\BibitemShut {NoStop}%
\bibitem [{\citenamefont {Hofstadter}(1976)}]{Hofstadter76}%
  \BibitemOpen
  \bibfield  {author} {\bibinfo {author} {\bibfnamefont {D.~R.}\ \bibnamefont
  {Hofstadter}},\ }\bibfield  {title} {\enquote {\bibinfo {title} {{Energy
  levels and wave functions of Bloch electrons in rational and irrational
  magnetic fields}},}\ }\href {\doibase 10.1103/PhysRevB.14.2239} {\bibfield
  {journal} {\bibinfo  {journal} {Phys. Rev.}\ }\textbf {\bibinfo {volume}
  {14}},\ \bibinfo {pages} {2239} (\bibinfo {year} {1976})}\BibitemShut
  {NoStop}%
\bibitem [{\citenamefont {Wilkinson}(1984)}]{Wilkinson84}%
  \BibitemOpen
  \bibfield  {author} {\bibinfo {author} {\bibfnamefont {M.}~\bibnamefont
  {Wilkinson}},\ }\bibfield  {title} {\enquote {\bibinfo {title} {{Critical
  Properties of Electron Eigenstates in Incommensurate Systems}},}\ }\href
  {\doibase 10.1098/rspa.1984.0016} {\bibfield  {journal} {\bibinfo  {journal}
  {Proc. R. Soc.}\ }\textbf {\bibinfo {volume} {391}},\ \bibinfo {pages} {305}
  (\bibinfo {year} {1984})}\BibitemShut {NoStop}%
\bibitem [{\citenamefont {Fuchs}\ and\ \citenamefont {Vidal}(2016)}]{Fuchs16}%
  \BibitemOpen
  \bibfield  {author} {\bibinfo {author} {\bibfnamefont {J.N.}\ \bibnamefont
  {Fuchs}}\ and\ \bibinfo {author} {\bibfnamefont {J.}~\bibnamefont {Vidal}},\
  }\bibfield  {title} {\enquote {\bibinfo {title} {{Hofstadter butterfly of a
  quasicrystal}},}\ }\href {\doibase 10.1103/PhysRevB.94.205437} {\bibfield
  {journal} {\bibinfo  {journal} {Phys. Rev. B}\ }\textbf {\bibinfo {volume}
  {94}},\ \bibinfo {pages} {205437} (\bibinfo {year} {2016})}\BibitemShut
  {NoStop}%
\bibitem [{\citenamefont {Duneau}\ and\ \citenamefont {Katz}(1985)}]{Duneau85}%
  \BibitemOpen
  \bibfield  {author} {\bibinfo {author} {\bibfnamefont {M.}~\bibnamefont
  {Duneau}}\ and\ \bibinfo {author} {\bibfnamefont {A.}~\bibnamefont {Katz}},\
  }\bibfield  {title} {\enquote {\bibinfo {title} {{Quasiperiodic Patterns}},}\
  }\href {\doibase 10.1103/PhysRevLett.54.2688} {\bibfield  {journal} {\bibinfo
   {journal} {Phys. Rev. Lett.}\ }\textbf {\bibinfo {volume} {54}},\ \bibinfo
  {pages} {2688} (\bibinfo {year} {1985})}\BibitemShut {NoStop}%
\bibitem [{\citenamefont {Kalugin}\ \emph {et~al.}()\citenamefont {Kalugin},
  \citenamefont {Kitaev},\ and\ \citenamefont {Levitov}}]{Kalugin85}%
  \BibitemOpen
  \bibfield  {author} {\bibinfo {author} {\bibfnamefont {P.~A.}\ \bibnamefont
  {Kalugin}}, \bibinfo {author} {\bibfnamefont {A.~Yu.}\ \bibnamefont
  {Kitaev}}, \ and\ \bibinfo {author} {\bibfnamefont {L.~S.}\ \bibnamefont
  {Levitov}},\ }\href@noop {} {\enquote {\bibinfo {title}
  {{Al$_{0.86}$Mn$_{0.14}$: a six-dimensional crystal}},}\ }\bibinfo {note}
  {\href{http://www.jetpletters.ac.ru/ps/1442/article_21941.shtml}{JETP Lett.
  {\bf 41}, 145 (1985)}}\BibitemShut {NoStop}%
\bibitem [{\citenamefont {Elser}(1986)}]{Elser85}%
  \BibitemOpen
  \bibfield  {author} {\bibinfo {author} {\bibfnamefont {V.}~\bibnamefont
  {Elser}},\ }\bibfield  {title} {\enquote {\bibinfo {title} {{The Diffraction
  Pattern of Projected Structures}},}\ }\href {\doibase
  10.1107/S0108767386099932} {\bibfield  {journal} {\bibinfo  {journal} {Acta
  Crystallogr., Sect. A: Found. Crystallogr.}\ }\textbf {\bibinfo {volume}
  {42}},\ \bibinfo {pages} {36} (\bibinfo {year} {1986})}\BibitemShut {NoStop}%
\bibitem [{\citenamefont {Vidal}\ and\ \citenamefont
  {Mosseri}(2001)}]{Vidal01}%
  \BibitemOpen
  \bibfield  {author} {\bibinfo {author} {\bibfnamefont {J.}~\bibnamefont
  {Vidal}}\ and\ \bibinfo {author} {\bibfnamefont {R.}~\bibnamefont
  {Mosseri}},\ }\bibfield  {title} {\enquote {\bibinfo {title} {{Generalized
  quasiperiodic Rauzy tilings}},}\ }\href {\doibase
  10.1088/0305-4470/34/18/317} {\bibfield  {journal} {\bibinfo  {journal} {J.
  Phys. A}\ }\textbf {\bibinfo {volume} {34}},\ \bibinfo {pages} {3927}
  (\bibinfo {year} {2001})}\BibitemShut {NoStop}%
\bibitem [{\citenamefont {Duneau}\ \emph {et~al.}(1989)\citenamefont {Duneau},
  \citenamefont {Mosseri},\ and\ \citenamefont {Oguey}}]{Duneau89}%
  \BibitemOpen
  \bibfield  {author} {\bibinfo {author} {\bibfnamefont {M.}~\bibnamefont
  {Duneau}}, \bibinfo {author} {\bibfnamefont {R.}~\bibnamefont {Mosseri}}, \
  and\ \bibinfo {author} {\bibfnamefont {C.}~\bibnamefont {Oguey}},\ }\bibfield
   {title} {\enquote {\bibinfo {title} {{Approximants of quasiperiodic
  structures generated by the inflation mapping}},}\ }\href {\doibase
  10.1088/0305-4470/22/21/017} {\bibfield  {journal} {\bibinfo  {journal} {J.
  Phys. A}\ }\textbf {\bibinfo {volume} {22}},\ \bibinfo {pages} {4549}
  (\bibinfo {year} {1989})}\BibitemShut {NoStop}%
\bibitem [{\citenamefont {Duneau}\ and\ \citenamefont {Audier}()}]{Duneau94}%
  \BibitemOpen
  \bibfield  {author} {\bibinfo {author} {\bibfnamefont {M.}~\bibnamefont
  {Duneau}}\ and\ \bibinfo {author} {\bibfnamefont {M.}~\bibnamefont
  {Audier}},\ }\href@noop {} {}\bibinfo {note} {\textit{Lectures on
  Quasicrystals}, edited by F. Hippert and D. Gratias (Les Editions de
  Physique, Les Ulis, 1994), p. 283}\BibitemShut {NoStop}%
\bibitem [{\citenamefont {Goldman}\ and\ \citenamefont
  {Kelton}(1993)}]{Goldman93}%
  \BibitemOpen
  \bibfield  {author} {\bibinfo {author} {\bibfnamefont {A.~I.}\ \bibnamefont
  {Goldman}}\ and\ \bibinfo {author} {\bibfnamefont {R.~F.}\ \bibnamefont
  {Kelton}},\ }\bibfield  {title} {\enquote {\bibinfo {title} {{Quasicrystals
  and crystalline approximants}},}\ }\href {\doibase 10.1103/RevModPhys.65.213}
  {\bibfield  {journal} {\bibinfo  {journal} {Rev. Mod. Phys.}\ }\textbf
  {\bibinfo {volume} {65}},\ \bibinfo {pages} {213} (\bibinfo {year}
  {1993})}\BibitemShut {NoStop}%
\bibitem [{\citenamefont {Vidal}\ and\ \citenamefont
  {Mosseri}(2004)}]{Vidal04}%
  \BibitemOpen
  \bibfield  {author} {\bibinfo {author} {\bibfnamefont {J.}~\bibnamefont
  {Vidal}}\ and\ \bibinfo {author} {\bibfnamefont {R.}~\bibnamefont
  {Mosseri}},\ }\bibfield  {title} {\enquote {\bibinfo {title} {{Quasiperiodic
  tilings under magnetic field}},}\ }\href {\doibase
  10.1016/j.jnoncrysol.2003.11.027} {\bibfield  {journal} {\bibinfo  {journal}
  {J. Non-Cryst. Solids}\ }\textbf {\bibinfo {volume} {334}},\ \bibinfo {pages}
  {130} (\bibinfo {year} {2004})}\BibitemShut {NoStop}%
\bibitem [{\citenamefont {Tran}\ \emph {et~al.}(2015)\citenamefont {Tran},
  \citenamefont {Dauphin}, \citenamefont {Goldman},\ and\ \citenamefont
  {Gaspard}}]{Tran15}%
  \BibitemOpen
  \bibfield  {author} {\bibinfo {author} {\bibfnamefont {D.-T.}\ \bibnamefont
  {Tran}}, \bibinfo {author} {\bibfnamefont {A.}~\bibnamefont {Dauphin}},
  \bibinfo {author} {\bibfnamefont {N.}~\bibnamefont {Goldman}}, \ and\
  \bibinfo {author} {\bibfnamefont {P.}~\bibnamefont {Gaspard}},\ }\bibfield
  {title} {\enquote {\bibinfo {title} {{Topological Hofstadter insulators in a
  two-dimensional quasicrystal}},}\ }\href {\doibase
  10.1103/PhysRevB.91.085125} {\bibfield  {journal} {\bibinfo  {journal} {Phys.
  Rev. B}\ }\textbf {\bibinfo {volume} {91}},\ \bibinfo {pages} {085125}
  (\bibinfo {year} {2015})}\BibitemShut {NoStop}%
\bibitem [{\citenamefont {Beenker}()}]{Beenker82}%
  \BibitemOpen
  \bibfield  {author} {\bibinfo {author} {\bibfnamefont {F.~P.~M.}\
  \bibnamefont {Beenker}},\ }\href@noop {} {\enquote {\bibinfo {title}
  {{Algebraic theory of non periodic tilings of the plane by two simple
  building blocks: a square and a rhombus}},}\ }\bibinfo {note}
  {\href{https://pure.tue.nl/ws/files/4269555/253166.pdf}{TH Report 82-WSK-04,
  (Technische Hogeschool, Eindhoven 1982)}}\BibitemShut {NoStop}%
\bibitem [{\citenamefont {Soret}()}]{Soret16}%
  \BibitemOpen
  \bibfield  {author} {\bibinfo {author} {\bibfnamefont {A.}~\bibnamefont
  {Soret}},\ }\href@noop {} {\enquote {\bibinfo {title} {{Quasicristaux sous
  champ magn\'etique}},}\ }\bibinfo {note} {Master's thesis, UPMC Paris,
  2016}\BibitemShut {NoStop}%
\bibitem [{\citenamefont {Penrose}(1974)}]{Penrose74}%
  \BibitemOpen
  \bibfield  {author} {\bibinfo {author} {\bibfnamefont {R.}~\bibnamefont
  {Penrose}},\ }\bibfield  {title} {\enquote {\bibinfo {title} {{The role of
  aesthetics in pure and applied mathematical research}},}\ }\href@noop {}
  {\bibfield  {journal} {\bibinfo  {journal} {Bull. Inst. Math. Appl.}\
  }\textbf {\bibinfo {volume} {10}},\ \bibinfo {pages} {266} (\bibinfo {year}
  {1974})}\BibitemShut {NoStop}%
\bibitem [{\citenamefont {Hatakeyama}\ and\ \citenamefont
  {Kamimura}(1989)}]{Hatakeyama89}%
  \BibitemOpen
  \bibfield  {author} {\bibinfo {author} {\bibfnamefont {T.}~\bibnamefont
  {Hatakeyama}}\ and\ \bibinfo {author} {\bibfnamefont {H.}~\bibnamefont
  {Kamimura}},\ }\bibfield  {title} {\enquote {\bibinfo {title} {{Fractal
  Nature of the Electronic Structure of a Penrose Tiling Lattice in a Magnetic
  Field}},}\ }\href {\doibase 10.1143/JPSJ.58.260} {\bibfield  {journal}
  {\bibinfo  {journal} {J. Phys. Soc. Jpn.}\ }\textbf {\bibinfo {volume}
  {58}},\ \bibinfo {pages} {260} (\bibinfo {year} {1989})}\BibitemShut
  {NoStop}%
\bibitem [{\citenamefont {Analytis}\ \emph {et~al.}(2005)\citenamefont
  {Analytis}, \citenamefont {Blundell},\ and\ \citenamefont
  {Ardavan}}]{Analytis05}%
  \BibitemOpen
  \bibfield  {author} {\bibinfo {author} {\bibfnamefont {J.~G.}\ \bibnamefont
  {Analytis}}, \bibinfo {author} {\bibfnamefont {S.~J.}\ \bibnamefont
  {Blundell}}, \ and\ \bibinfo {author} {\bibfnamefont {A.}~\bibnamefont
  {Ardavan}},\ }\bibfield  {title} {\enquote {\bibinfo {title} {{Magnetic
  oscillations, disorder and the Hofstadter butterfly in finite systems}},}\
  }\href {\doibase 10.1016/j.synthmet.2005.07.068} {\bibfield  {journal}
  {\bibinfo  {journal} {Synthetic Metals}\ }\textbf {\bibinfo {volume} {154}},\
  \bibinfo {pages} {265} (\bibinfo {year} {2005})}\BibitemShut {NoStop}%
\bibitem [{\citenamefont {Wegner}(1983)}]{Wegner83}%
  \BibitemOpen
  \bibfield  {author} {\bibinfo {author} {\bibfnamefont {F.}~\bibnamefont
  {Wegner}},\ }\bibfield  {title} {\enquote {\bibinfo {title} {{Exact density
  of states for lowest Landau level in white noise potential superfield
  representation for interacting systems}},}\ }\href {\doibase
  10.1007/BF01319209} {\bibfield  {journal} {\bibinfo  {journal} {Z. Phys. B}\
  }\textbf {\bibinfo {volume} {51}},\ \bibinfo {pages} {279} (\bibinfo {year}
  {1983})}\BibitemShut {NoStop}%
\bibitem [{\citenamefont {Kittel}()}]{Kittel}%
  \BibitemOpen
  \bibfield  {author} {\bibinfo {author} {\bibfnamefont {C.}~\bibnamefont
  {Kittel}},\ }\href@noop {} {\enquote {\bibinfo {title} {{\it Introduction to
  Solid State Physics}},}\ }\bibinfo {note} {3rd ed. (Wiley, New York,
  1967)}\BibitemShut {NoStop}%
\bibitem [{\citenamefont {Gambaudo}\ and\ \citenamefont
  {Vignolo}(2014)}]{Gambaudo14}%
  \BibitemOpen
  \bibfield  {author} {\bibinfo {author} {\bibfnamefont {J.~M.}\ \bibnamefont
  {Gambaudo}}\ and\ \bibinfo {author} {\bibfnamefont {P.}~\bibnamefont
  {Vignolo}},\ }\bibfield  {title} {\enquote {\bibinfo {title} {{Brillouin zone
  labelling for quasicrystals}},}\ }\href {\doibase
  10.1088/1367-2630/16/4/043013} {\bibfield  {journal} {\bibinfo  {journal}
  {New J. Phys.}\ }\textbf {\bibinfo {volume} {16}},\ \bibinfo {pages} {043013}
  (\bibinfo {year} {2014})}\BibitemShut {NoStop}%
\bibitem [{\citenamefont {Harper}(1955)}]{Harper55}%
  \BibitemOpen
  \bibfield  {author} {\bibinfo {author} {\bibfnamefont {P.~G.}\ \bibnamefont
  {Harper}},\ }\bibfield  {title} {\enquote {\bibinfo {title} {{Single Band
  Motion of Conduction Electrons in a Uniform Magnetic Field}},}\ }\href
  {\doibase 10.1088/0370-1298/68/10/304} {\bibfield  {journal} {\bibinfo
  {journal} {Proc. Phys. Soc. A}\ }\textbf {\bibinfo {volume} {68}},\ \bibinfo
  {pages} {874} (\bibinfo {year} {1955})}\BibitemShut {NoStop}%
\bibitem [{\citenamefont {Landau}\ and\ \citenamefont {Lifshitz}()}]{LL3}%
  \BibitemOpen
  \bibfield  {author} {\bibinfo {author} {\bibfnamefont {L.D.}\ \bibnamefont
  {Landau}}\ and\ \bibinfo {author} {\bibfnamefont {E.~M.}\ \bibnamefont
  {Lifshitz}},\ }\href@noop {} {\enquote {\bibinfo {title} {{\it Quantum
  Mechanics: Non-relativistic theory}},}\ }\bibinfo {note} {2nd ed.
  (Butterworth-Heinemann, Oxford, 1965) Vol. 3, Sec. 50, Problem 3}\BibitemShut
  {NoStop}%
\bibitem [{\citenamefont {Grimm}\ and\ \citenamefont {Schreiber}()}]{Grimm03}%
  \BibitemOpen
  \bibfield  {author} {\bibinfo {author} {\bibfnamefont {U.}~\bibnamefont
  {Grimm}}\ and\ \bibinfo {author} {\bibfnamefont {M.}~\bibnamefont
  {Schreiber}},\ }\href@noop {} {\enquote {\bibinfo {title} {{Energy spectra
  and eigenstates of quasiperiodic tight-binding Hamiltonians}},}\ }\bibinfo
  {note} {In \textit{Quasicrystals: Structure and physical properties}, edited
  by H.-R. Trebin (Wiley-VCH, Weinheim, 2003), pp. 210-235}\BibitemShut
  {NoStop}%
\bibitem [{\citenamefont {Mac\'e}\ \emph {et~al.}(2017)\citenamefont {Mac\'e},
  \citenamefont {Jagannathan}, \citenamefont {Kalugin}, \citenamefont
  {Mosseri},\ and\ \citenamefont {Pi\'echon}}]{Mace17}%
  \BibitemOpen
  \bibfield  {author} {\bibinfo {author} {\bibfnamefont {N.}~\bibnamefont
  {Mac\'e}}, \bibinfo {author} {\bibfnamefont {A.}~\bibnamefont {Jagannathan}},
  \bibinfo {author} {\bibfnamefont {P.}~\bibnamefont {Kalugin}}, \bibinfo
  {author} {\bibfnamefont {R.}~\bibnamefont {Mosseri}}, \ and\ \bibinfo
  {author} {\bibfnamefont {F.}~\bibnamefont {Pi\'echon}},\ }\bibfield  {title}
  {\enquote {\bibinfo {title} {{Critical eigenstates and their properties in
  one- and two-dimensional quasicrystals}},}\ }\href {\doibase
  10.1103/PhysRevB.96.045138} {\bibfield  {journal} {\bibinfo  {journal} {Phys.
  Rev. B}\ }\textbf {\bibinfo {volume} {96}},\ \bibinfo {pages} {045138}
  (\bibinfo {year} {2017})}\BibitemShut {NoStop}%
\bibitem [{\citenamefont {Kalugin}\ and\ \citenamefont
  {Katz}(2014)}]{Kalugin14}%
  \BibitemOpen
  \bibfield  {author} {\bibinfo {author} {\bibfnamefont {P.}~\bibnamefont
  {Kalugin}}\ and\ \bibinfo {author} {\bibfnamefont {A.}~\bibnamefont {Katz}},\
  }\bibfield  {title} {\enquote {\bibinfo {title} {{Electrons in deterministic
  quasicrystalline potentials and hidden conserved quantities}},}\ }\href
  {\doibase 10.1088/1751-8113/47/31/315206} {\bibfield  {journal} {\bibinfo
  {journal} {J. Phys. A: Math. Theor.}\ }\textbf {\bibinfo {volume} {47}},\
  \bibinfo {pages} {315206} (\bibinfo {year} {2014})}\BibitemShut {NoStop}%
\bibitem [{\citenamefont {Sutherland}(1987)}]{Sutherland87}%
  \BibitemOpen
  \bibfield  {author} {\bibinfo {author} {\bibfnamefont {B.}~\bibnamefont
  {Sutherland}},\ }\bibfield  {title} {\enquote {\bibinfo {title} {{Critical
  electronic wave functions on quasiperiodic lattices: Exact calculation of
  fractal measures}},}\ }\href {\doibase 10.1103/PhysRevB.35.9529} {\bibfield
  {journal} {\bibinfo  {journal} {Phys. Rev. B}\ }\textbf {\bibinfo {volume}
  {35}},\ \bibinfo {pages} {9529} (\bibinfo {year} {1987})}\BibitemShut
  {NoStop}%
\end{thebibliography}

%

\end{document}